\documentclass[fleqn,usenatbib]{mnras}
\usepackage{newtxtext,newtxmath}
\usepackage[T1]{fontenc}
\usepackage{ae,aecompl}

\usepackage{subfig,graphicx,caption}
\usepackage{amsmath}	    

\graphicspath{{./plots/}}

\title[Exotic Image Formation in Cluster Lenses]
{Exotic Image Formation in Strong Gravitational Lensing by Clusters 
of Galaxies -- II: Uncertainties}

\author[A. K. Meena et al.]{
Ashish Kumar Meena$^{1}$\thanks{E-mail: ashishmeena766@gmail.com},
Agniva Ghosh$^{2}$,
Jasjeet S. Bagla$^{1}$\thanks{E-mail: jasjeet@iisermohali.ac.in}
and Liliya L. R. Williams$^{2}$
\\
\\
$^{1}$Indian Institute of Science Education and Research Mohali,
Knowledge City, Sector 81, SAS Nagar, Punjab 140306,
India
\\
$^{2}$School of Physics $\&$ Astronomy, University of Minnesota, 116 
Church Street SE, Minneapolis, MN 55455, USA
\\}

\pubyear{2021}

\begin{document}
\label{firstpage}
\pagerange{\pageref{firstpage}--\pageref{lastpage}}
\maketitle

\begin{abstract}

Due to the finite amount of observational data, the best-fit 
parameters corresponding to the reconstructed cluster mass 
have uncertainties. 
In turn, these uncertainties affect the inferences made from 
these mass models. 
Following our earlier work,  we have studied the effect of 
such uncertainties on the singularity maps in simulated and 
actual galaxy clusters. 
The mass models for both simulated and real clusters have been 
constructed using \textsc{grale}. 
The final best-fit mass models created using 
\textsc{grale} give the simplest singularity maps and a lower limit on 
the number of point singularities that a lens has to offer. 
The simple nature of these singularity maps also puts a lower 
limit on the number of three image (tangential and radial) 
arcs that a cluster lens has.
Hence, we estimate the number of galaxy sources giving rise 
to the three image arcs, which can be observed with the 
\textit{James Webb Space Telescope} (JWST). 
We find that we expect to observe at least 20-30 tangential 
and 5-10 radial three-image arcs in the Hubble Frontier Fields 
cluster lenses with the JWST.

\end{abstract}

\begin{keywords}
gravitational lensing: strong -- galaxies: clusters: individual (Abell 370, 
Abell 2744, Abell S1063, MACS J0416.1-2403, MACSJ0717.5+3745, MACS J1149.5+2223)
\end{keywords}

\section{Introduction}
\label{sec: Introduction}

Galaxy clusters as strong lenses offer a unique probe to the matter 
distribution as lensing does not discriminate between the visible and 
dark matter distribution~\citep[e.g.,][]{1992ARA&A..30..311B, 2011A&ARv..19...47K}.
At the same time, by magnifying distant sources (which would have 
otherwise remained unobserved), it also opens up the possibility of 
studying the high-redshift Universe. 
Observation of these cluster lenses led us to the detection of multiply 
imaged supernovae \citep{2015Sci...347.1123K, 2016ApJ...819L...8K}, 
which can help us in constraining the Hubble constant.
On the other hand, the observation of highly magnified stars
\citep{2018NatAs...2..334K} can help us in probing the mass-function 
of intra-cluster medium~\citep{2018ApJ...857...25D} 
and observing the pop III stars~\citep{2018ApJS..234...41W}.
Due to the high magnification provided by lensing, the detection of 
high-redshift {($z>7$) galaxies \citep[e.g.,][]{2008ApJ...678..647B, 
2013ApJ...762...32C, 2015Natur.519..327W} can help in determination of
the galaxy luminosity function~\citep[e.g.,][]{2018MNRAS.479.5184A}.

In order to use galaxy clusters as a probe, we need to model their mass 
distributions.
However, as the data from the observation is limited, one cannot model
the mass distribution of these clusters with arbitrary precision and 
resolution.
As a result, to reconstruct the lens mass distribution, different 
groups start with different sets of prior.
For examples, light trace mass~\citep[LTM,][]{2005ApJ...621...53B, 
2009MNRAS.396.1985Z} 
parametric mass reconstruction assumes that the mass distribution 
in the cluster lenses follows the light distribution.
On the other hand, the non-parametric \citep[free-form,][]{2007MNRAS.380.1729L} 
mass reconstruction methods do not rely on any preliminary information 
related to the mass model and only take into account the strong and weak
\citep{2020MNRAS.494.3253L} lensing data.
Hybrid mass reconstruction methods take input from both parametric
and non-parametric approaches~\citep{2014MNRAS.437.2642S}.
Due to the finite amount of data and different set of priors, it is
possible that these different methods give different results when 
applied on the same cluster lens~\citep[e.g.,][]{2009ApJ...707L.163S,
2009ApJ...703L.132Z}.
Hence, it is very important to compare these different techniques
in case of simulated as well as real lenses to improve these methods
and for more robust predictions~\citep[e.g.,][]{2017MNRAS.472.3177M, 
2017MNRAS.465.1030P, 2015ApJ...811...70R, 2018ApJ...863...60R, 2020MNRAS.494.4771R}.
Each of these methods results into the best-fit mass model parameters
and uncertainties associated with them.
One avenue to observe the effect of these uncertainties is the 
cluster lens magnification maps. 
For the best-fit lens mass model, one will have a certain area in 
the lens plane that gives higher magnification than a threshold value. 
However, once we account for the statistical uncertainties, the area 
with a magnification greater than the threshold area is not a unique 
number; instead, it will be denoted by a range. 
As a result, various observational estimates (like the number of highly 
magnified galaxies) are also subjected to these uncertainties.

For a source to be highly magnified, the source must lie close to the
caustic in the source plane and the corresponding images are formed
near the corresponding critical line in the lens plane.
The magnification factor by which the source is magnified also depends 
on the source size: smaller the source, larger the magnification factor, 
and vice versa. 
This is why one can have a magnification factor of thousands for 
lensed stellar sources~\citep{2018NatAs...2..334K}, whereas the maximum 
magnification factor for galaxy sources can only be in hundreds.
These caustics and critical lines are the singularities of the lens
mapping and are of two different types: stable and unstable.
Fold and cusp are stable singularities as they are present for all
source redshift, whereas beak-to-beak, umbilics, and swallowtails are
unstable singularities as they only occur for specific source redshifts
for a given lens system.
So far, the number of known image formation near unstable singularities is
very small: one near hyperbolic umbilic~\citep{2009MNRAS.399....2O, 
2010MNRAS.408.1916Z}, handful near swallowtail~\citep[e.g.,][]{1998MNRAS.294..734A,
2010A&A...524A..94S},
and none near elliptic umbilic.
With the upcoming observing facilities like: 
Euclid:~\citep{2009arXiv0912.0914L}, 
James Webb Space Telescope:~\citep[JWST,][]{2006SSRv..123..485G}, 
Nancy Grace Roman Space Telescope:~\citep[WFIRST,][]{2019arXiv190205569A}, 
Vera Rubin Observatory:~\citep[LSST,][]{2019ApJ...873..111I}, 
the number of strongly lensed system is expected to increase by more 
than an order of magnitude, and it is likely that we will encounter image 
formations near these unstable singularities.
Hence, it is timely to explore the properties of these unstable
singularities in detail.

In our previous work, we have described the method to locate these
unstable singularities for a given lens model~\citep[][hereafter 
MB20]{2020MNRAS.492.3294M} and applied this method in the case of actual 
cluster lenses~\citep[][hereafter MB21]{2021MNRAS.503.2097M}.
The final output of this method is a singularity map containing all the 
point singularities and $A_3$-lines.
Since these $A_3$-lines correspond to cusps in the source plane, 
they mark the high-magnification regions in the lens where image formation 
near cusps will occur (three or more lensed images lying near to each other).
The point singularities depend on the second and higher-order derivatives of 
the lens potential. 
Hence, they are very sensitive to the presence of small/intermediate-scale 
structures and the variations in the lens parameters. 
As a result, the singularity maps corresponding to the parametric and 
non-parametric mass models can be very different for the same cluster lens
As pointed out in MB21, parametric mass models give  significantly larger 
number of point singularities due to the small scale structures compared to 
non-parametric mass models.

In our current work, we study the effect of the statistical uncertainties 
associated with the reconstructed lens mass model parameters on the 
singularity maps and the higher-order singularity cross-section.
In order to study the effect of mass model uncertainties on the
singularity maps, we have considered two simulated galaxy cluster lenses, 
Irtysh I and II from \citet[][hereafter GWL20]{2020MNRAS.494.3998G} 
and all of the six galaxy cluster lenses from the \textit{Hubble Frontier
Fields (HFF) survey}. 
The lens mass reconstruction of all of these cluster lenses has been
done using \textsc{grale}\footnote{\url{https://research.edm.uhasselt.be/jori/grale2/}}
\citep{2007MNRAS.380.1729L}.
We choose the \textsc{grale} mass models as the corresponding singularity maps are 
simplest and provide a lower limit on the point singularity cross-section (MB21).
Further, as the \textsc{grale} mass model does not make any assumption about density 
profiles of substructure, the reconstruction is independent of the nature of dark matter.  
Due to the simple nature, we can construct singularity maps for a larger
region of the lens plane without introducing spurious point singularities. 
Apart from that, different \textsc{grale} runs for a cluster lens give independent 
mass maps. 
Hence, there is no correlation between different mass maps reconstructed 
for a lens using \textsc{grale} that may not be true for the parametric mass models.
For simulated clusters, Irtysh I and II, we construct the singularity
maps for the original mass models, for the individual runs (with 150,
500, 1000 lensed images) and for the final best-fit mass model.
For HFF clusters, we construct the singularity maps for the individual runs 
and for the best-fit mass model obtained by the averaging of the individual runs.

As we know that the $A_3$-lines in the singularity maps correspond to the 
cusps in the source plane.
Hence, a singularity map also gives the information about the number
of cusps formed in the source plane at a given source redshift.
By drawing the critical lines for a given source redshift overlaid on the 
singularity map, one can calculate the number of cusps in the source plane 
by counting the points where a critical line and corresponding $A_3$-line cut 
each other.
As the \textsc{grale} best-fit mass models provide the simplest singularity 
maps, they also give the lower limit on the cusp cross-section.
Hence, it is worthwhile to calculate the lower limit of the cusp cross-section
in the HFF clusters using the \textsc{grale} best-fit mass models.
Following that, we also estimate the (tangential and radial) arc
cross-section in case of both simulated and HFF cluster lenses.
The source galaxy luminosity function has been taken 
from~\citet[][hereafter C18]{2018MNRAS.474.2352C}.
In C18, authors estimate the number of galaxies expected in the deep 
galaxy surveys with the JWST in different the filters
for different exposure time. 
Although the expected galaxy population in any of the JWST filter can 
be used in our study, we only considered the expected galaxy population
in one JWST filter, F200W, with an exposure time of $10^4$ seconds
(please see C18 for more details).

This paper is organized as follows. In \S\ref{sec: singularities},
we briefly revisit the basics of the singularities in the strong lensing.
In \S\ref{sec: simulated}, we present our results for simulated Irtysh
clusters.
The corresponding singularity maps are discussed in 
\S\ref{ssec: simulated singularity maps}, followed by 
the discussion of the redshift distribution of singularities 
in \S\ref{ssec: simulated redshift distribution}.
The results for the HFF clusters are presented in \S\ref{sec: hff}.
The corresponding singularity maps, redshift distribution of singularities,
and the arc cross-section are discussed in
\S\ref{ssec: HFF singularity maps},
\S\ref{ssec: HFF redshift distribution},
and \S\ref{ssec: HFF arc cross section}, respectively.
Summary and conclusions are presented in \S\ref{sec: conclusions}.
The cosmological parameters used in this work to calculate the various 
quantities are: ${\rm H}_0=70\:{\rm kms}^{-1}{\rm Mpc}^{-1},\:
\Omega_\Lambda=0.7,\: \Omega_m=0.3$.

\section{Singularities in Strong Lensing}
\label{sec: singularities}

\begin{figure*}
  \includegraphics[width=\textwidth,height=8.0cm,width=8.0cm]{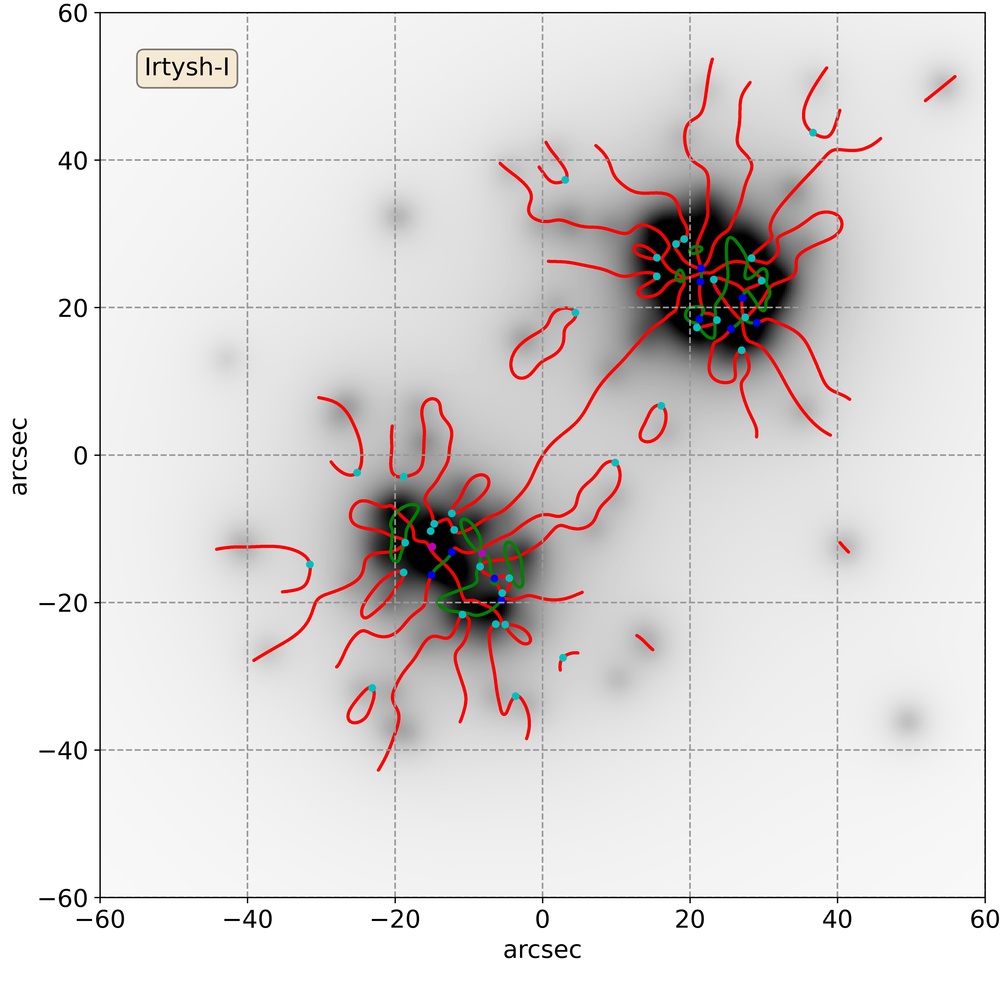}
  \includegraphics[width=\textwidth,height=8.0cm,width=8.0cm]{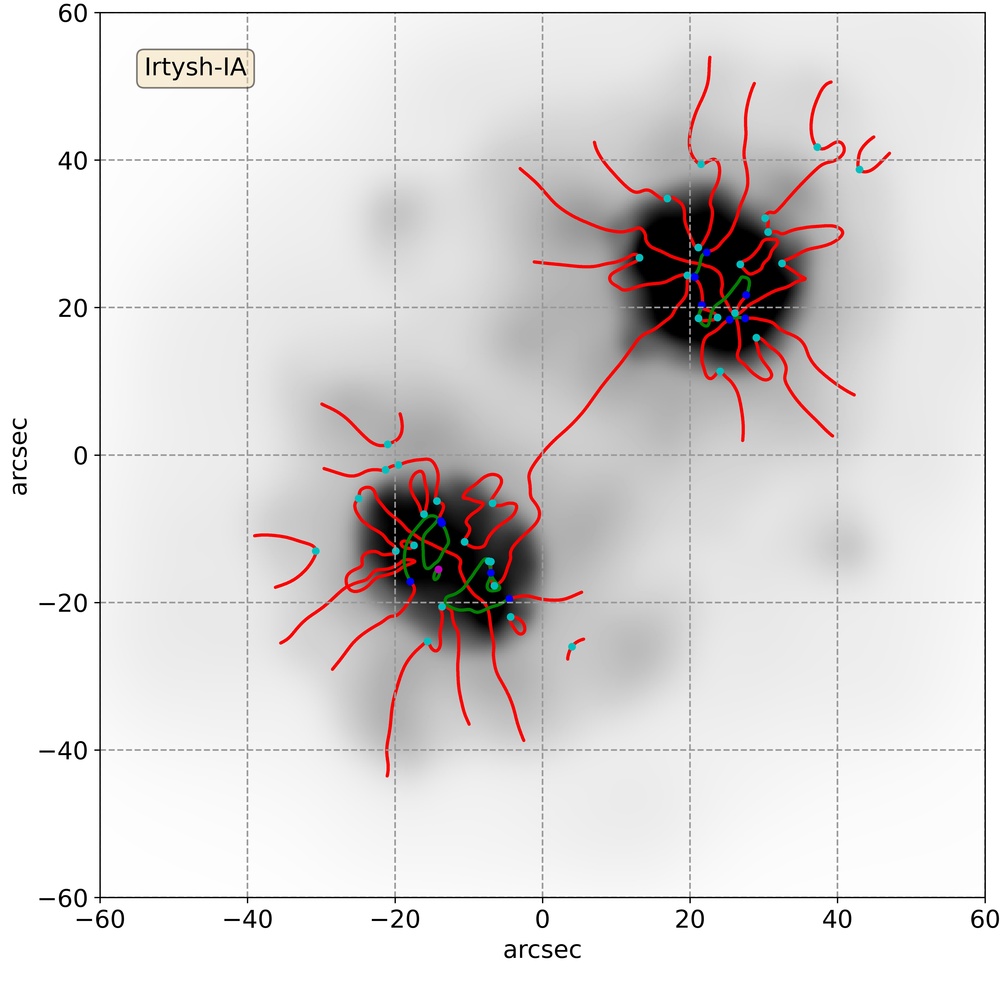}
  \includegraphics[width=\textwidth,height=8.0cm,width=8.0cm]{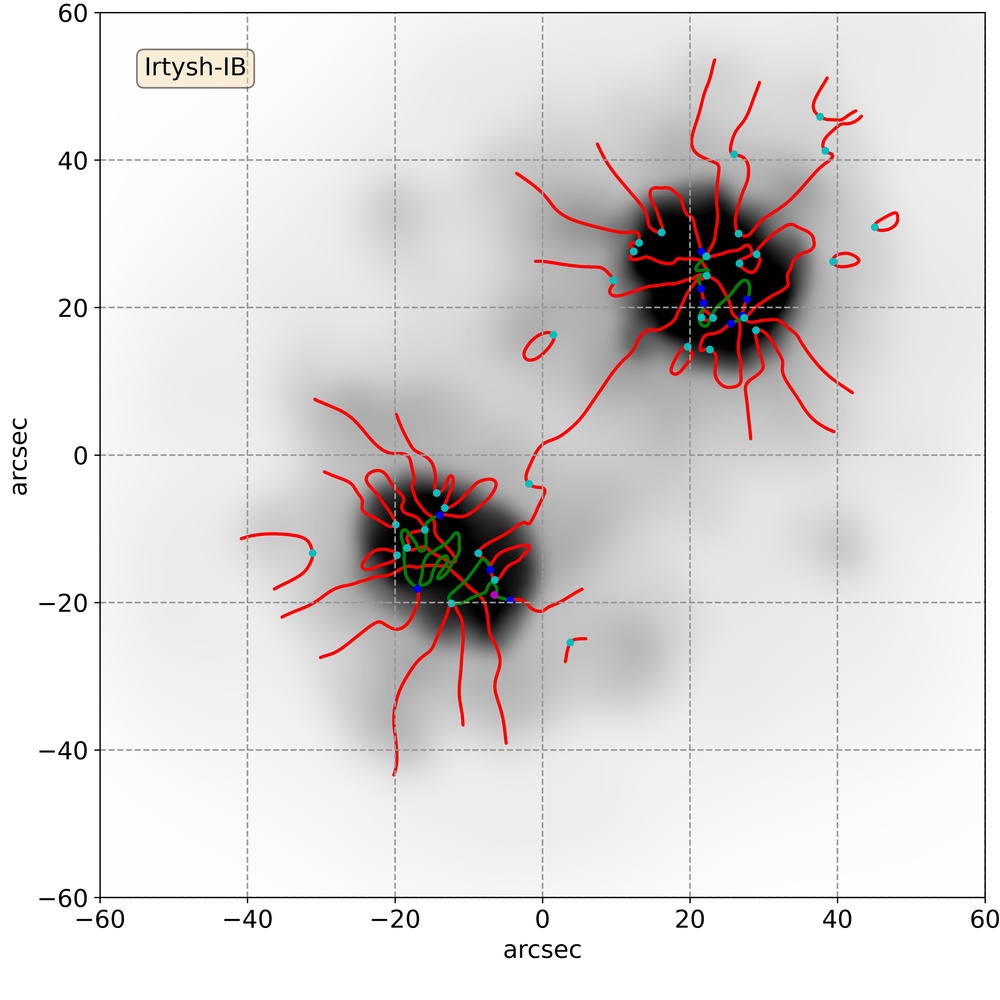}
  \includegraphics[width=\textwidth,height=8.0cm,width=8.0cm]{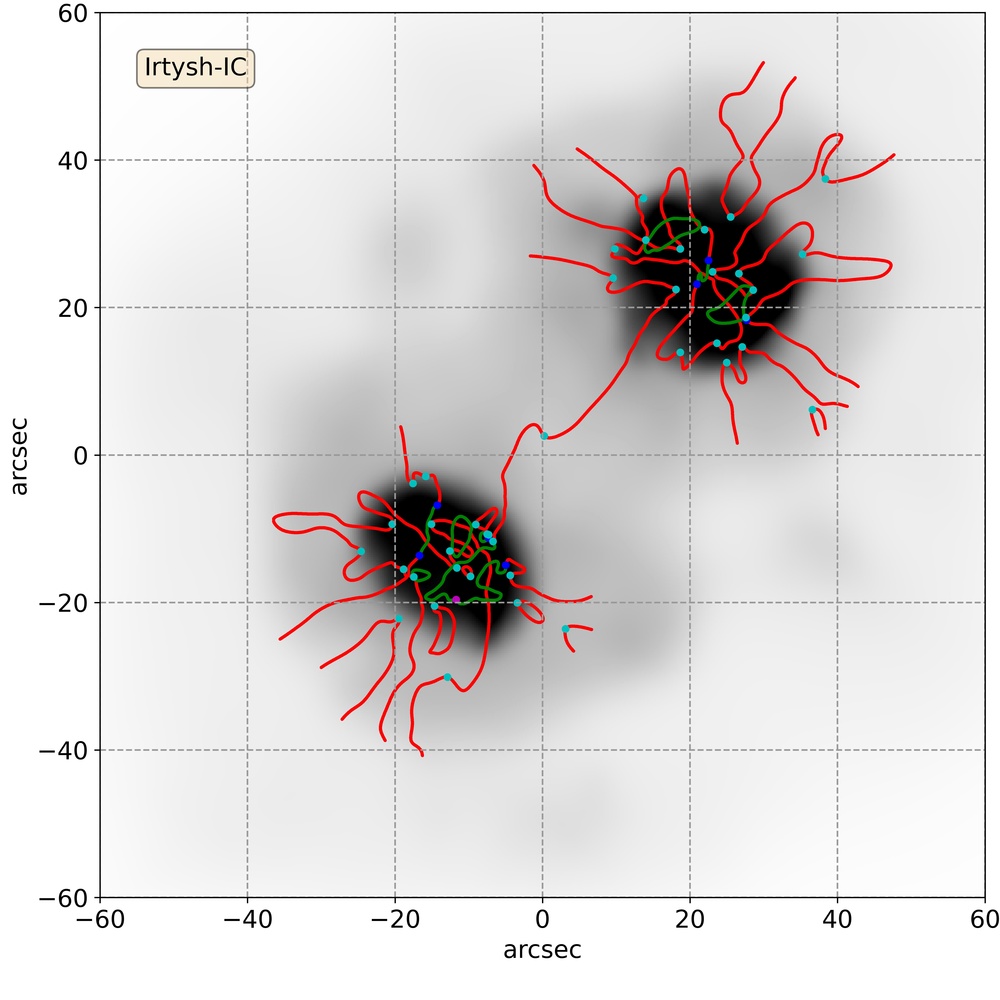}
  \caption{Singularity maps for original and reconstructed Irtysh I mass models:
  The top-left panel represents the singularity map corresponding to the original
  Irtysh I mass models. The top-right, bottom-left, bottom-right panels represent
  the singularity maps for Irtysh IA, IB, IC, respectively. In each panel, the red
  and green lines represent the $A_3$-lines corresponding to the $\alpha$ and $\beta$
  eigenvalues of the deformation tensor. The blue points represent the location of
  (hyperbolic and elliptic) umbilics. The cyan and magenta points represent the
  swallowtail singularities corresponding to the $\alpha$ and $\beta$ eigenvalues 
  of the deformation tensor. In each panel, the background is the corresponding 
  normalized mass distribution in the lens plane.}
  \label{fig: irtysh I singularity}
\end{figure*}

\begin{figure*}
  \includegraphics[width=\textwidth,height=8.0cm,width=8.0cm]{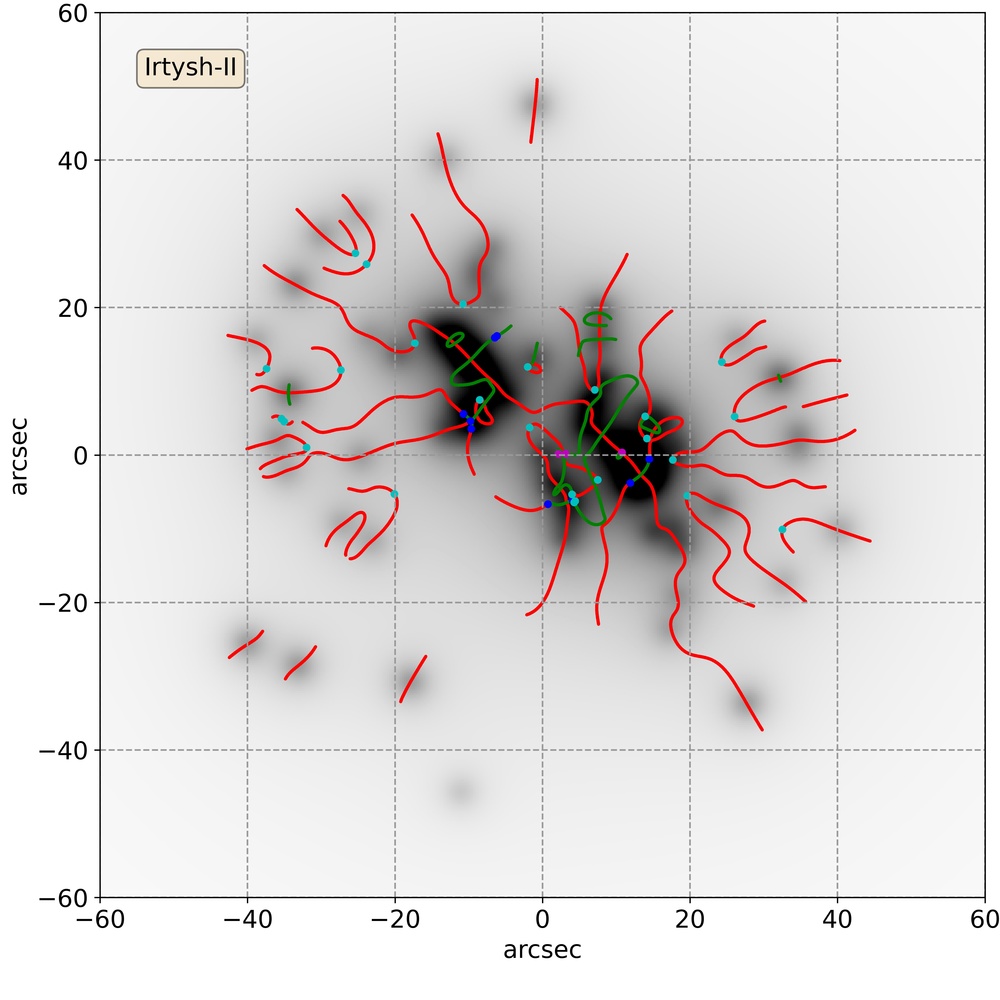}
  \includegraphics[width=\textwidth,height=8.0cm,width=8.0cm]{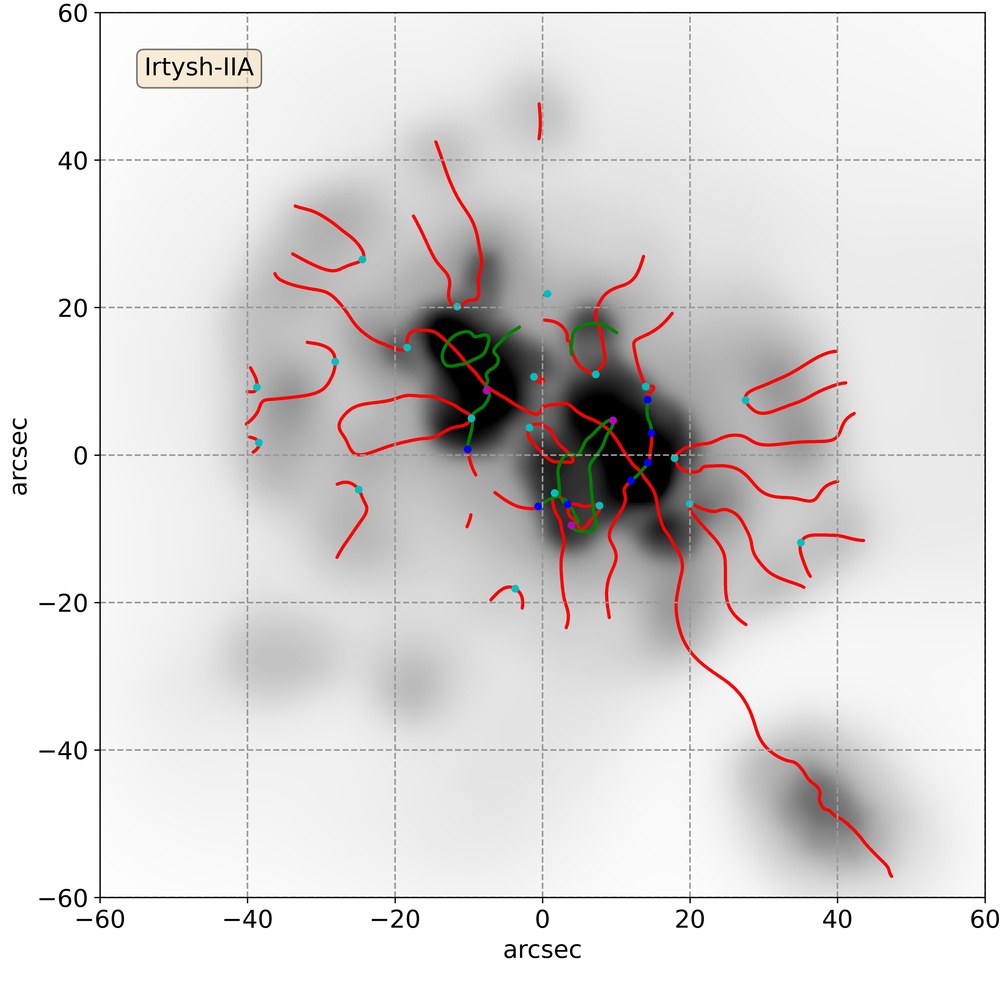}
  \includegraphics[width=\textwidth,height=8.0cm,width=8.0cm]{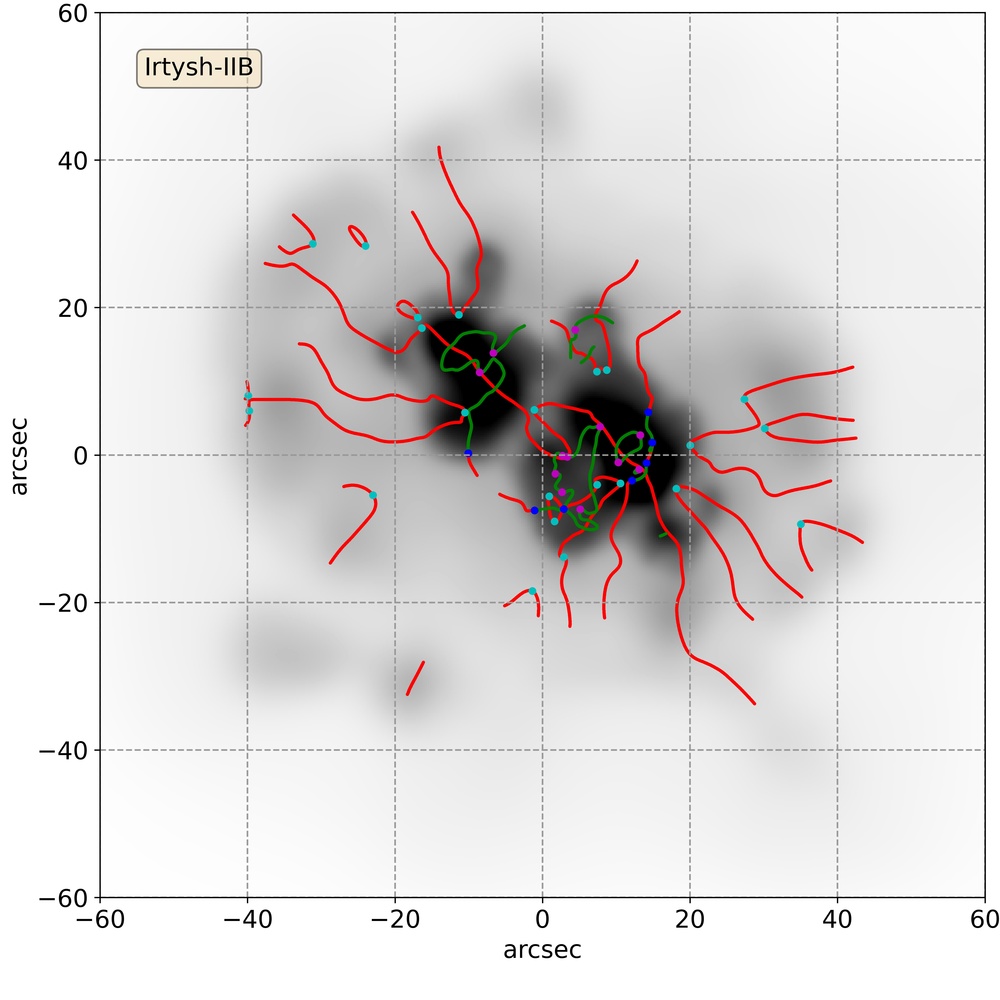}
  \includegraphics[width=\textwidth,height=8.0cm,width=8.0cm]{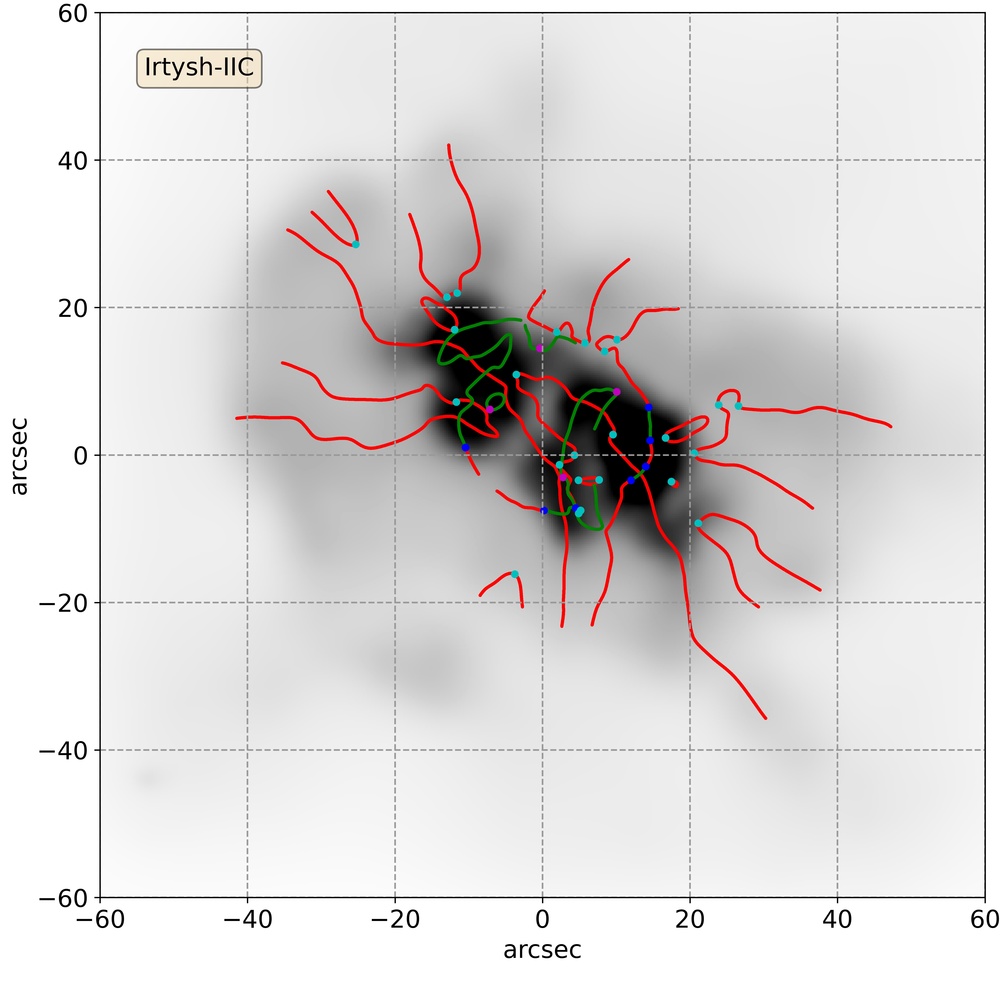}
  \caption{Singularity maps for original and reconstructed Irtysh II mass models:
  The top-left panel represents the singularity map corresponding to the original
  Irtysh II mass models. The top-right, bottom-left, bottom-right panels represent
  the singularity maps for Irtysh IIA, IIB, IIC, respectively. In each panel, the red
  and green lines represent the $A_3$-lines corresponding to the $\alpha$ and $\beta$
  eigenvalues of the deformation tensor. The blue points represent the location of
  (hyperbolic and elliptic) umbilics. The cyan and magenta points represent the
  swallowtail singularities corresponding to the $\alpha$ and $\beta$ eigenvalues 
  of the deformation tensor. In each panel, the background is the corresponding 
  normalized mass distribution in the lens plane.}
  \label{fig: irtysh II singularity}
\end{figure*}

\begin{figure*}
  \includegraphics[width=\textwidth,height=7.0cm,width=8.0cm]
  {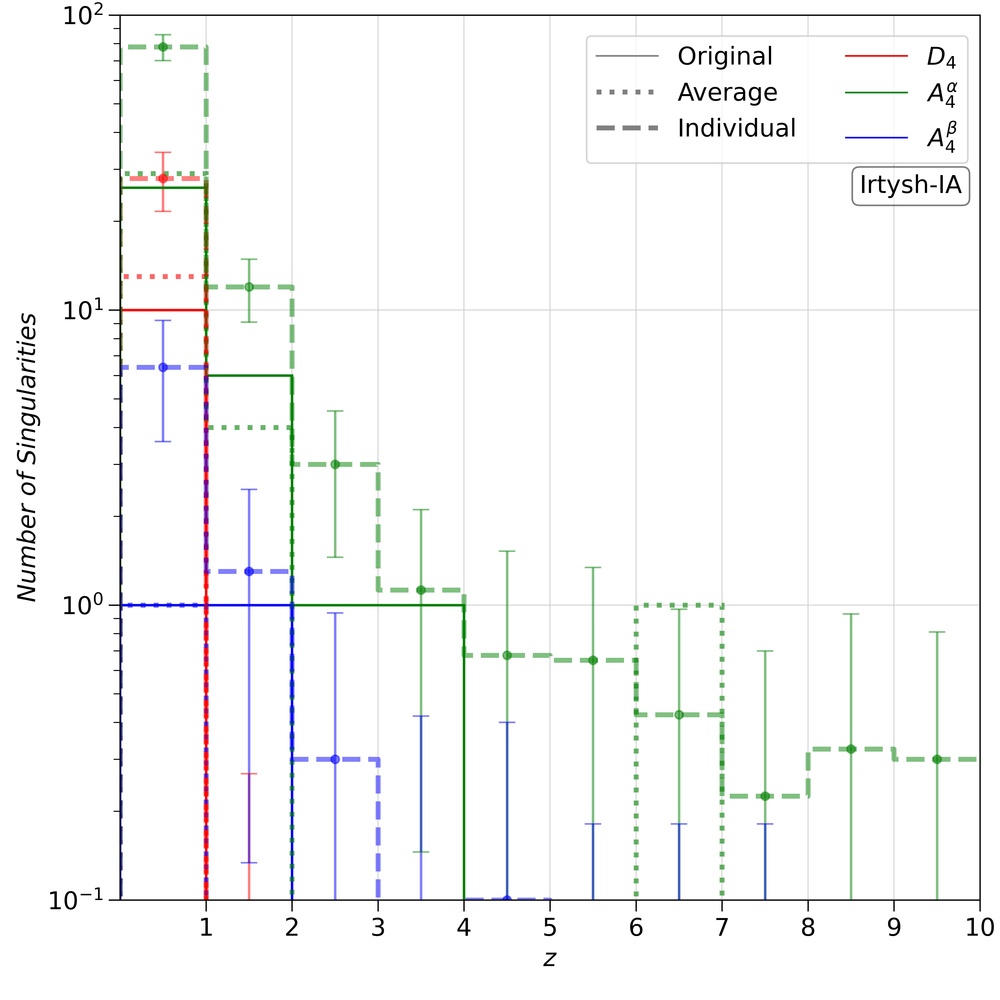}
  \includegraphics[width=\textwidth,height=7.0cm,width=8.0cm]
  {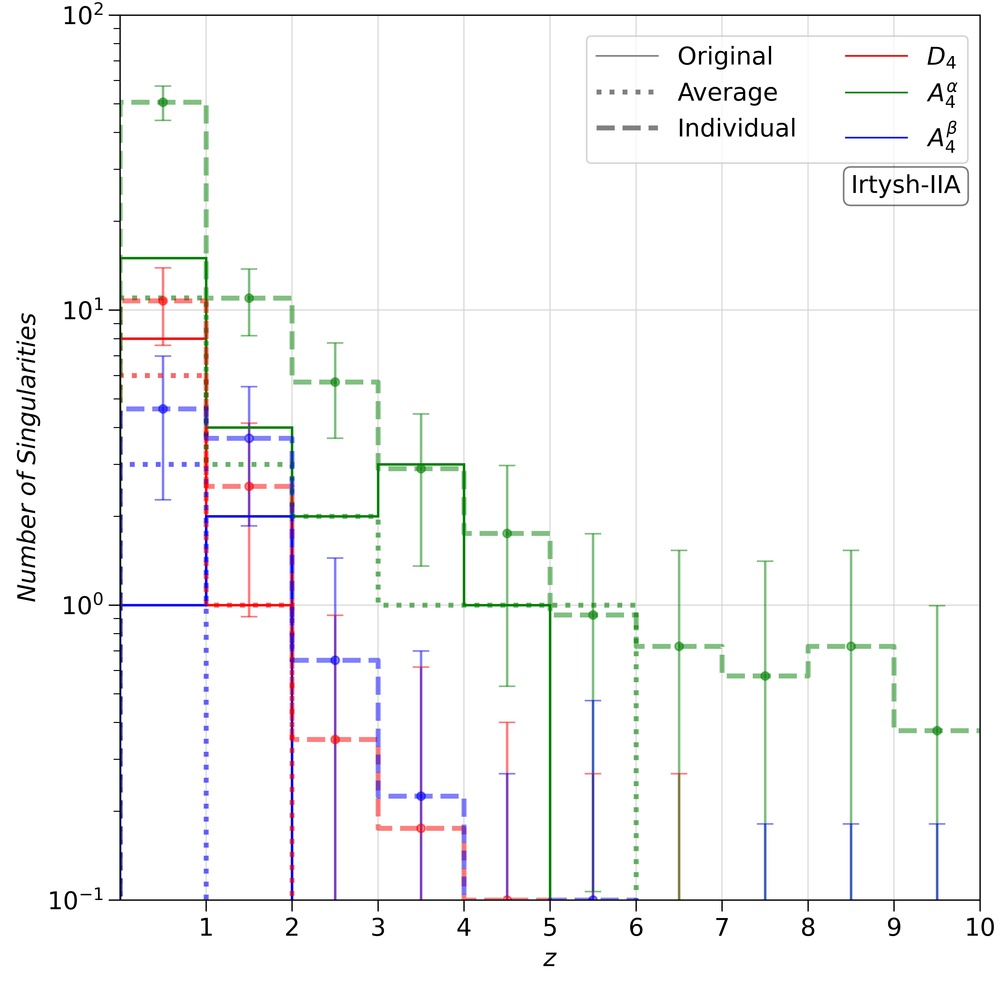}
  \includegraphics[width=\textwidth,height=7.0cm,width=8.0cm]
  {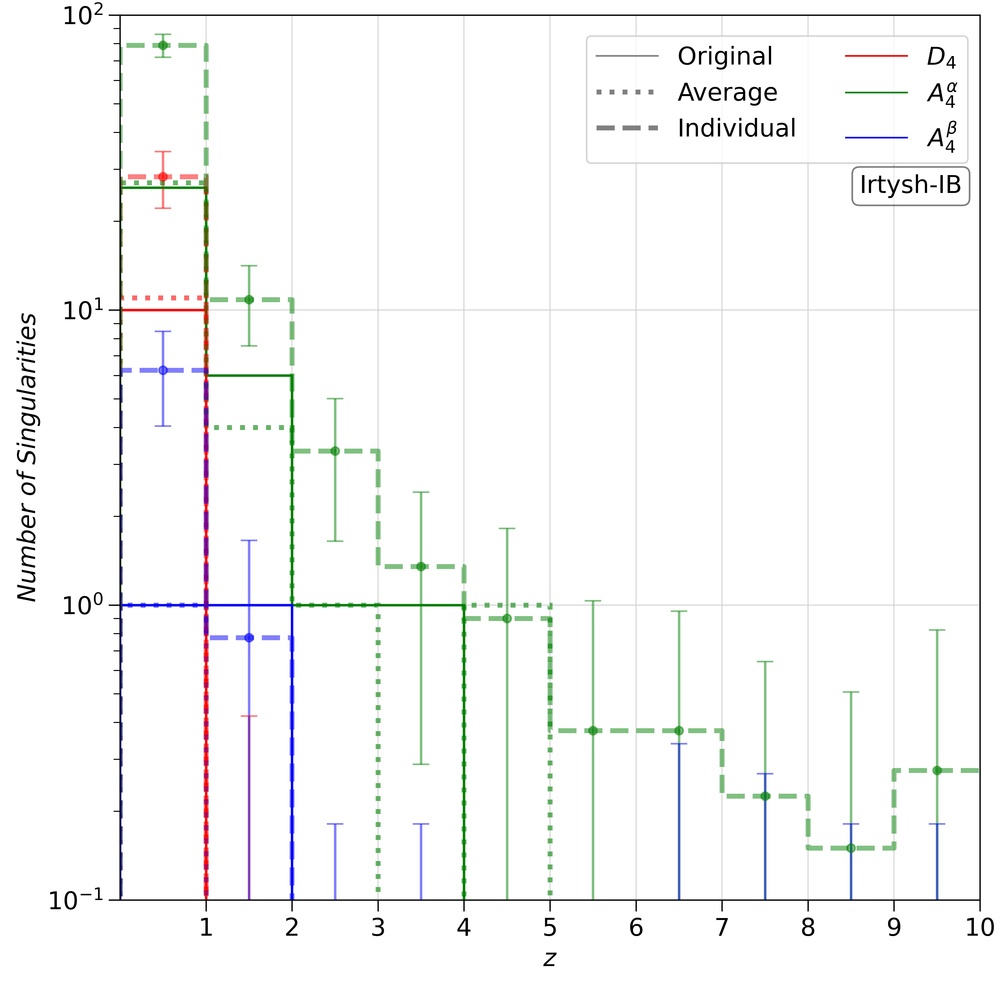}
  \includegraphics[width=\textwidth,height=7.0cm,width=8.0cm]
  {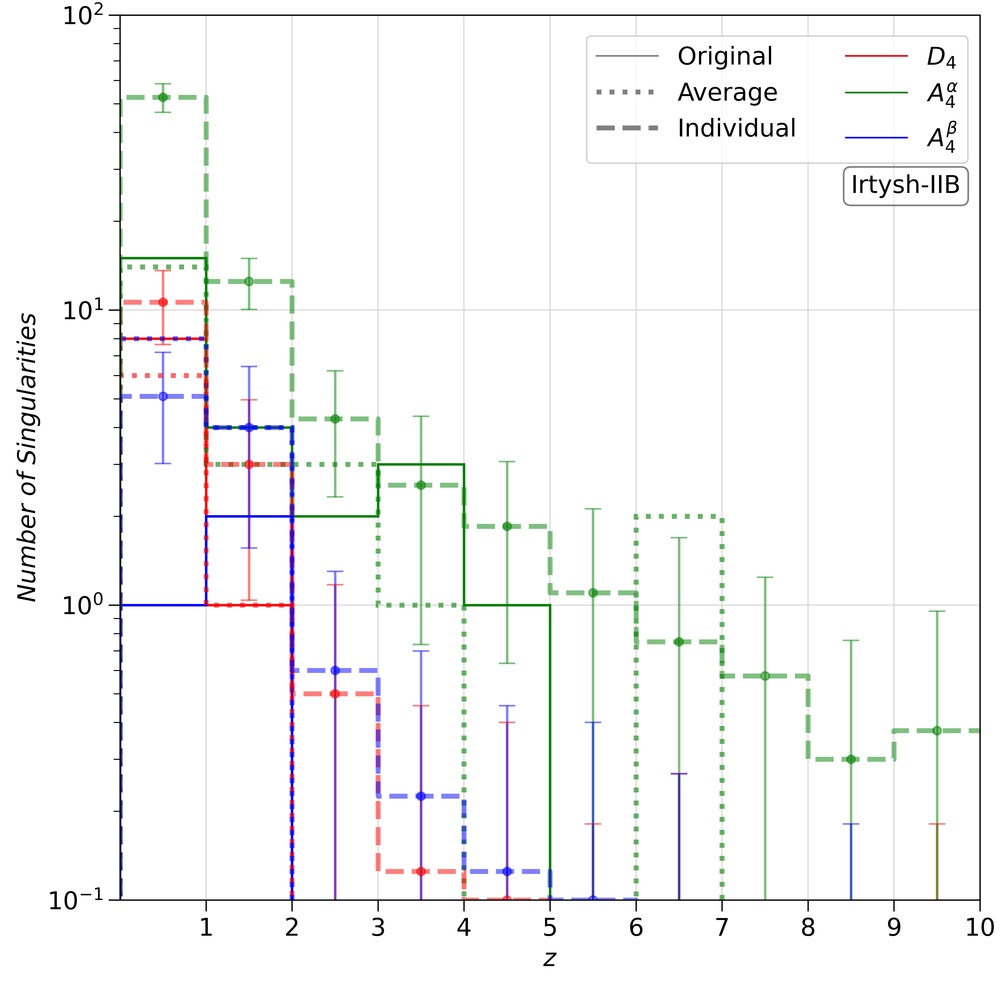}
  \includegraphics[width=\textwidth,height=7.0cm,width=8.0cm]
  {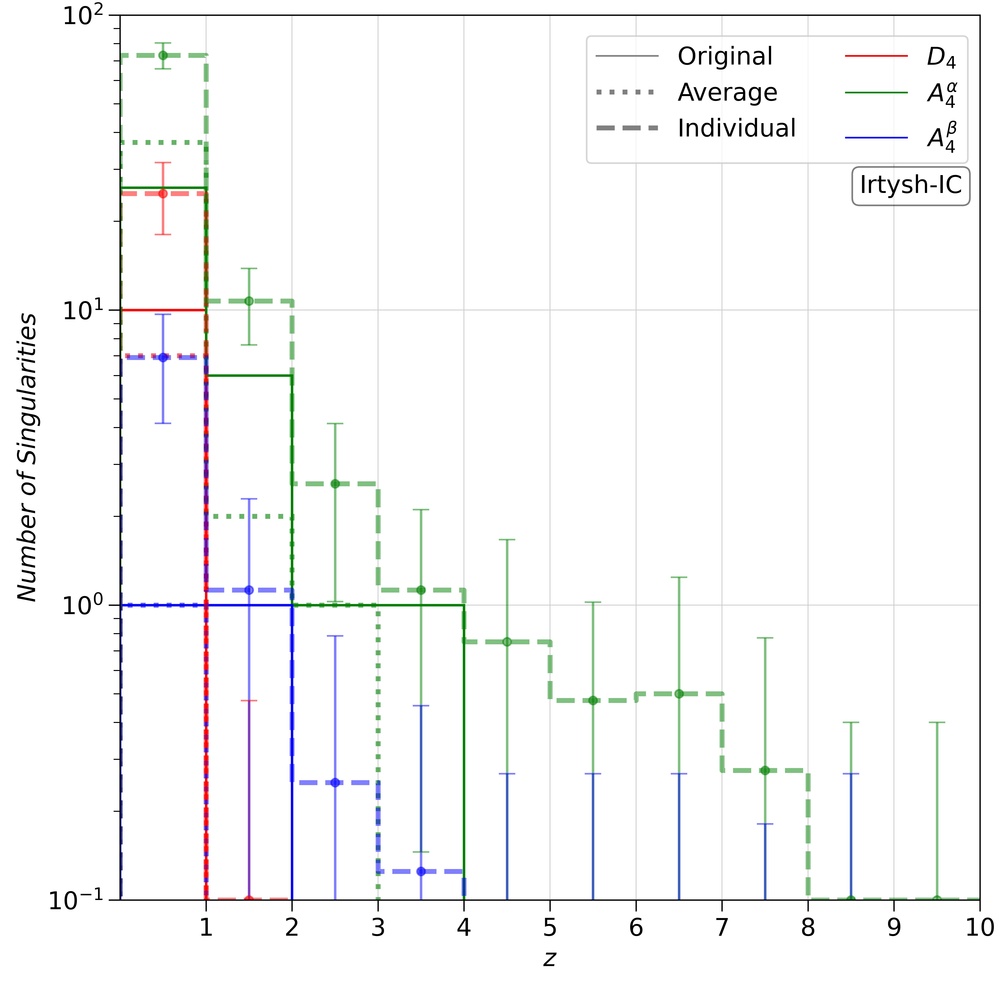}
  \includegraphics[width=\textwidth,height=7.0cm,width=8.0cm]
  {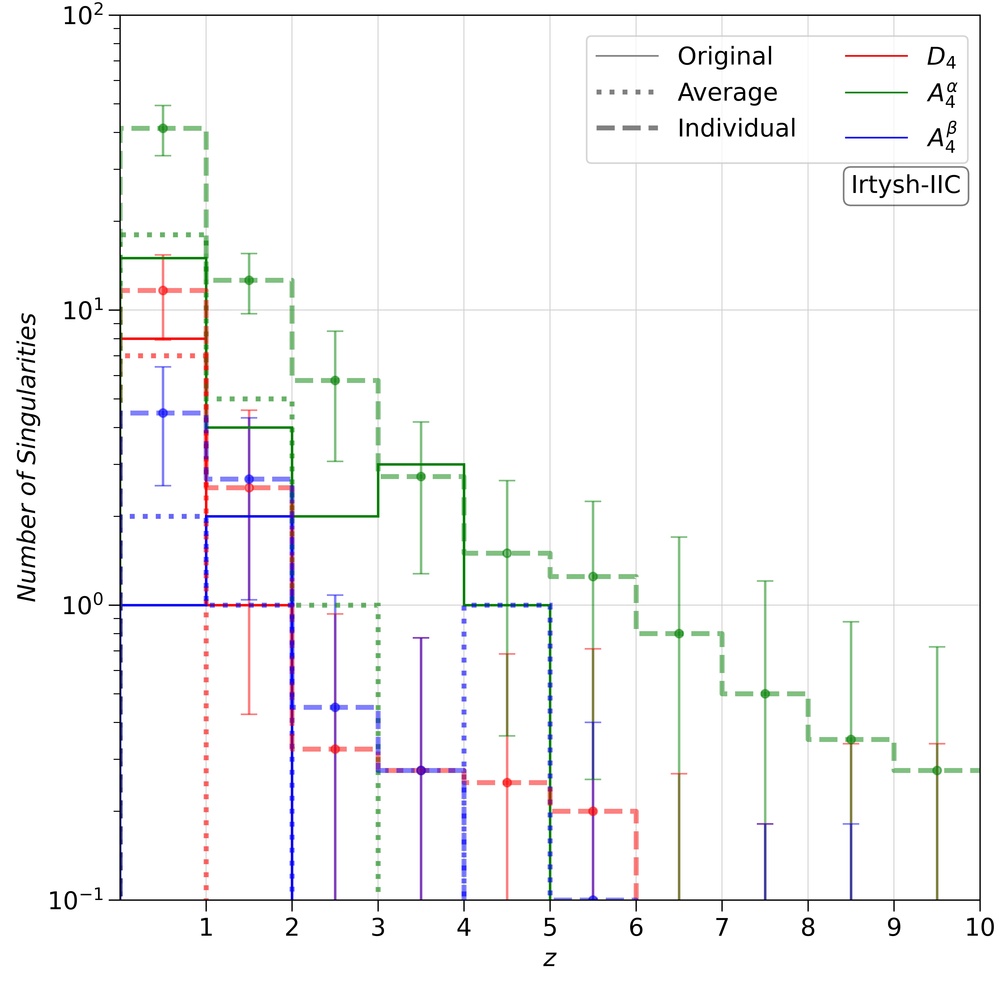}
  \caption{Number of singularities as a function of redshift in Irtysh I and II:
  In each panel of left (right) column, the thin lines represent the number of 
  singularities corresponding to the original Irtysh I (II). The thick dotted 
  lines represent the number of singularities corresponding to the final averaged 
  mass models of Irtysh I and II (the corresponding cluster name is written in each panel).
  The thick dashed lines show the average number of singularity in individual mass
  models for Irtysh I and II. The error bars represent the corresponding one sigma 
  scatter. The red lines represent the distribution of (hyperbolic$+$elliptic) umbilics.
  The green and blue lines represent the distribution of swallowtail singularities
  corresponding to $\alpha$ and $\beta$ eigenvalues of the deformation tensor.}
  \label{fig: Irtysh histogram}
\end{figure*}

In this section, we briefly review the basics of singularities in
strong gravitational lensing to set the notions and notations.
For a more detailed discussion about singularities, we encourage 
the reader to see into MB20, and~\citet{1992grle.book.....S}.

For a given strong lens system, magnification factor in the lens (image) 
plane is given as,
\begin{equation}
  \mu\left(\mathbf{x}\right)  =
  \frac{1}{\left(1-a\alpha\right)\left(1-a\beta\right)}, 
  \label{eq:(magnification)}
\end{equation}
where $\alpha$ and $\beta$ are the eigenvalues of the deformation 
tensor $\psi_{ij}$, with $\psi$ being the projected lens potential 
and subscript $ij$ represents its partial derivatives with 
respect to the image plane coordinates. 
$a=D_{\rm ds}/D_{\rm s}$ is the distance ratio, where $D_{\rm ds}$ 
and $D_{\rm s}$ represent the angular diameter distance  
from the lens to the source and from the observer to the source, respectively.
The points in the lens plane where $\mu^{-1} = 0$ are known as the
singular points of the lens mapping (see MB20 for more details).
The location of these singular points in the lens plane defines the 
critical curves and the corresponding points in the source plane 
are known as the caustics.
The critical curves in the lens plane are smooth closed curves, and
trace regions with high magnification, whereas their counterpart, 
caustic curves in the source are closed but not necessarily smooth.
These caustics in the source plane are made of smooth segments 
(known as folds) meeting with each other at the cusp points.

For a given strong lens system, fold and cusp are the only stable 
singularity of the lens mapping as they are present for all possible 
source redshifts.
The evolution of caustic structure in the source plane with source 
redshift mainly depicts the formation, destruction, or exchange of 
cusps between radial and tangential caustics.
Other higher-order singularities (also known as point singularities) 
like swallowtail, purse, and pyramid are only present for specific 
source redshifts and are sensitive to the lens parameters.
A small change in the lens system parameters can result in
the complete disappearance of a point singularity. 
These point singularities, along with the $A_3$-lines constitute
the so called \textit{singularity map}.
$A_3$-lines form the backbone of the singularity map and denote the
points in the lens plane which correspond to the cusps in the 
source plane for all possible source redshifts.
Different point singularities are associated with the formation, 
destruction, or exchange of cusp between radial and tangential 
caustics, hence, they also satisfy the $A_3$-line condition along 
with additional criteria.
$A_3$-lines are given by the condition $n_\lambda.\nabla_x\lambda = 0$,
where $\lambda$ denotes the eigenvalue of the deformation tensor, 
and $n_\lambda$ is the corresponding eigenvector.
The swallowtail singularity denotes the points in the lens plane 
where the eigenvector $n_\lambda$ is tangential to the corresponding 
$A_3$-line, and at a swallowtail singularity two extra cusps appear
in the source plane.
The hyperbolic and elliptic umbilics mark the points in the lens 
plane where both eigenvalues of the deformation tensor are equal 
to each other, i.e., zero shear points.
We encourage the reader to see MB20 for more details about the 
point singularities and the corresponding characteristic image formations.

As these point singularities depend on the second and high-order 
derivatives of the lens potential, the effect of variation of lens 
parameters can be seen clearly in the corresponding singularity map.
As a result, the singularity map is also very sensitive to the lens mass 
reconstruction method, as shown in MB21.
In MB21, one can see that the cluster mass models reconstructed 
using non-parametric method \textsc{grale} give a significantly lower number of 
point singularities compared to the parametric mass models for the 
same cluster, hence, putting the lower limit on the point singularity 
cross-section.
The singularity maps corresponding to the non-parametric mass models 
also show very simple $A_3$-line structures compared to the singularity
map corresponding to the parametric mass models.
Hence, the three image (tangential and radial) arc cross-section in 
parametric and non-parametric mass models is also expected to show 
significant differences.

\section{Simulated Clusters}
\label{sec: simulated}

\begin{figure*}
  \includegraphics[width=\textwidth,height=8.0cm,width=8.0cm]{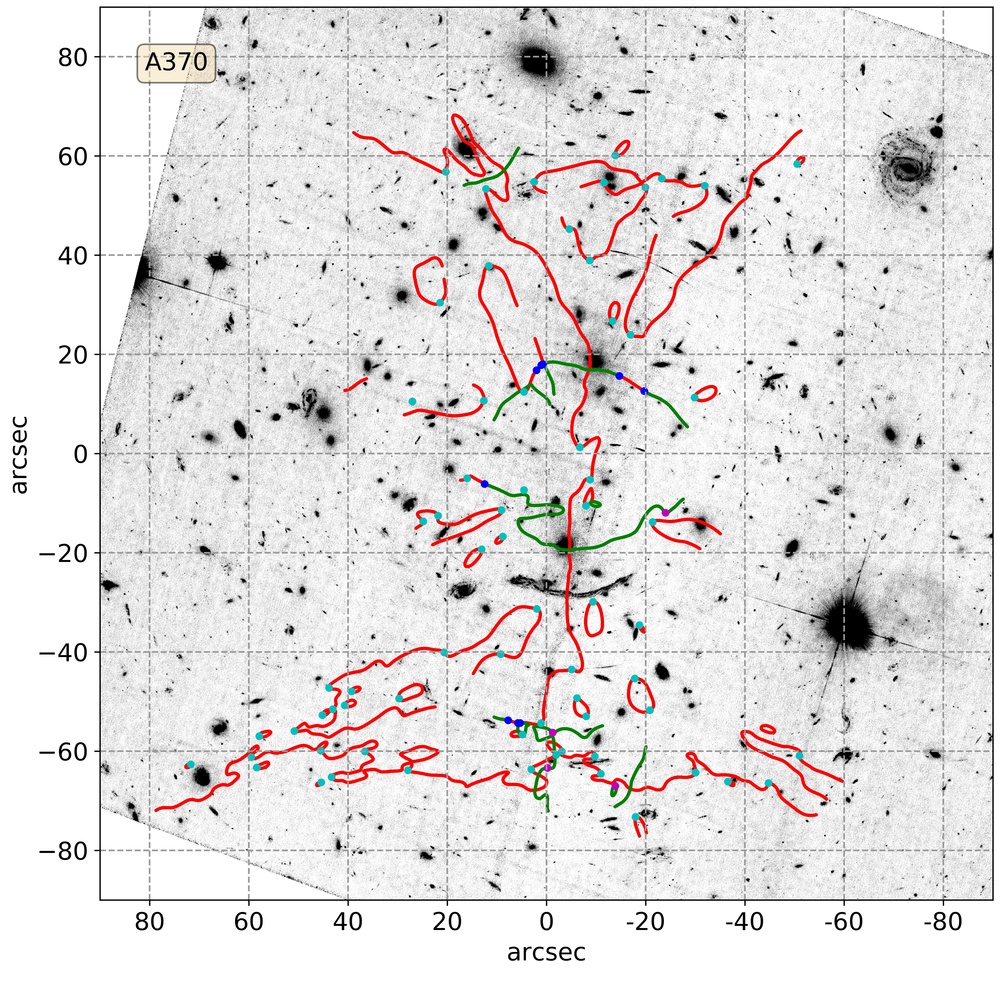}
  \includegraphics[width=\textwidth,height=8.0cm,width=8.0cm]{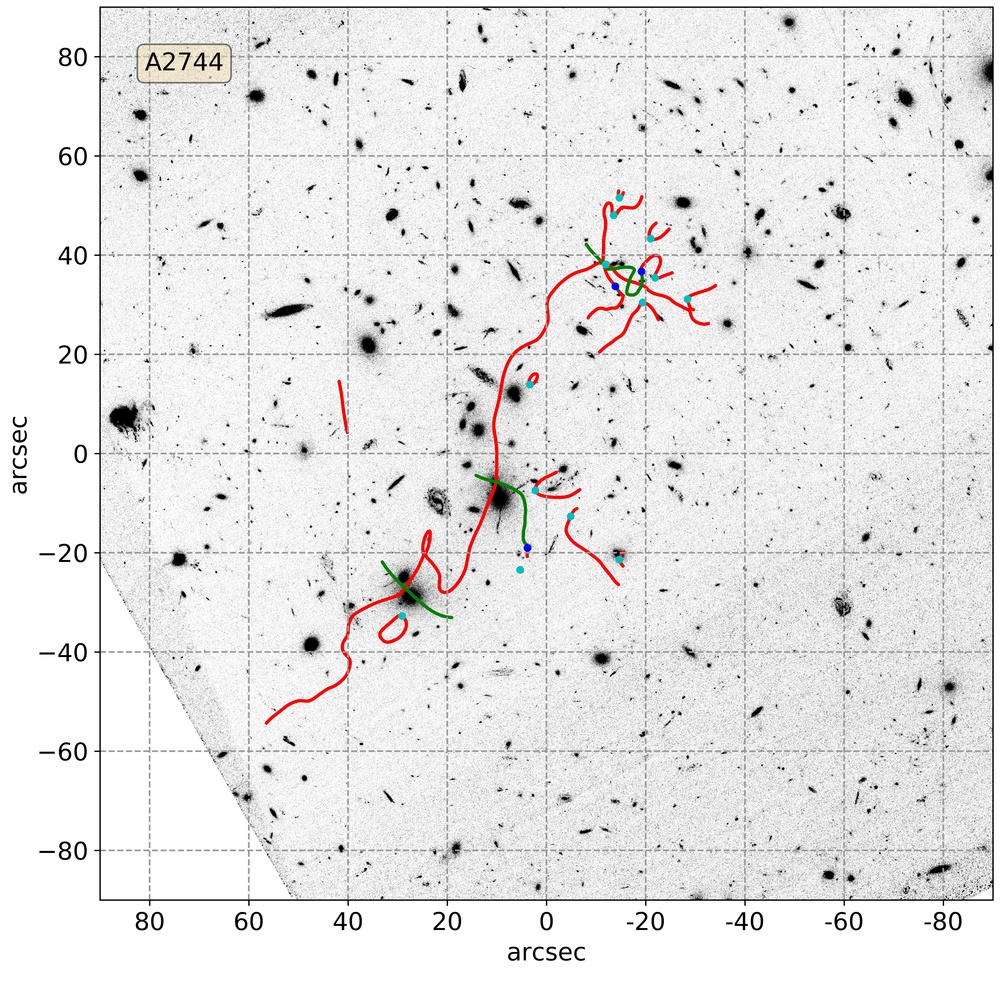}
  \includegraphics[width=\textwidth,height=8.0cm,width=8.0cm]{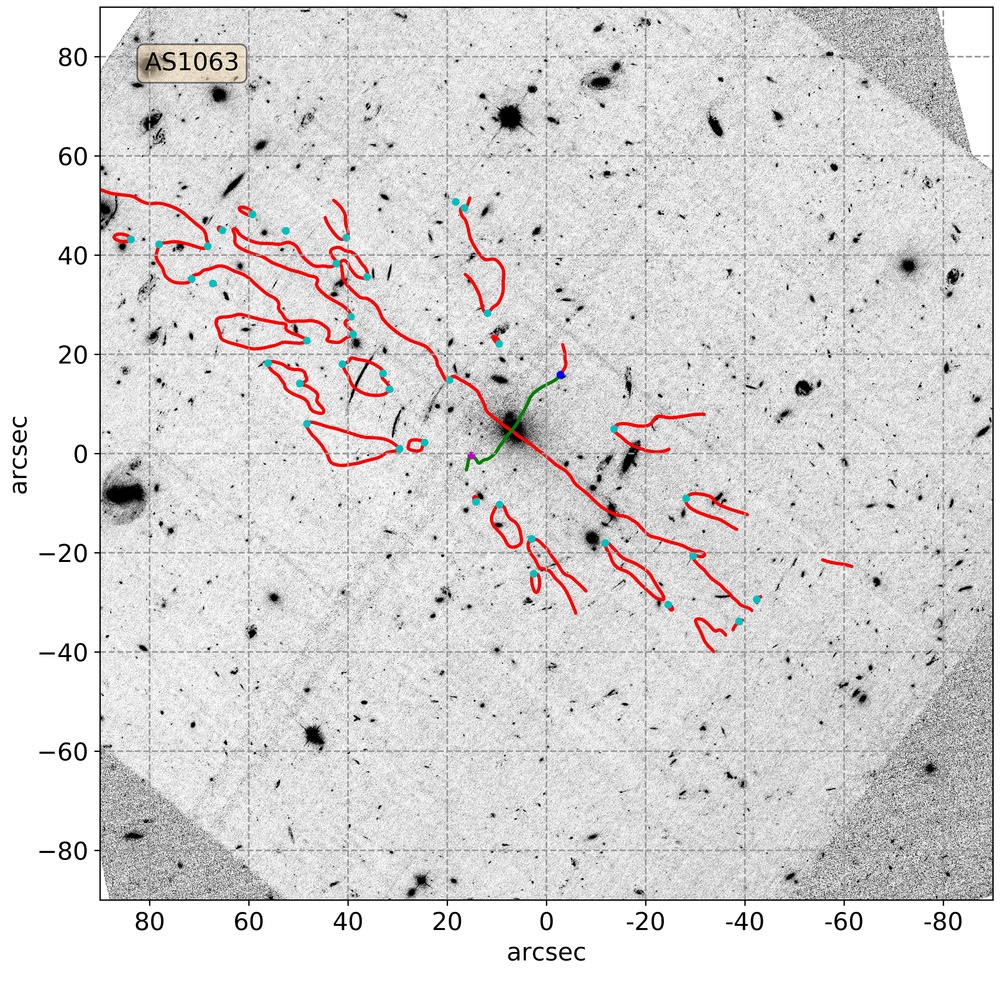}
  \includegraphics[width=\textwidth,height=8.0cm,width=8.0cm]{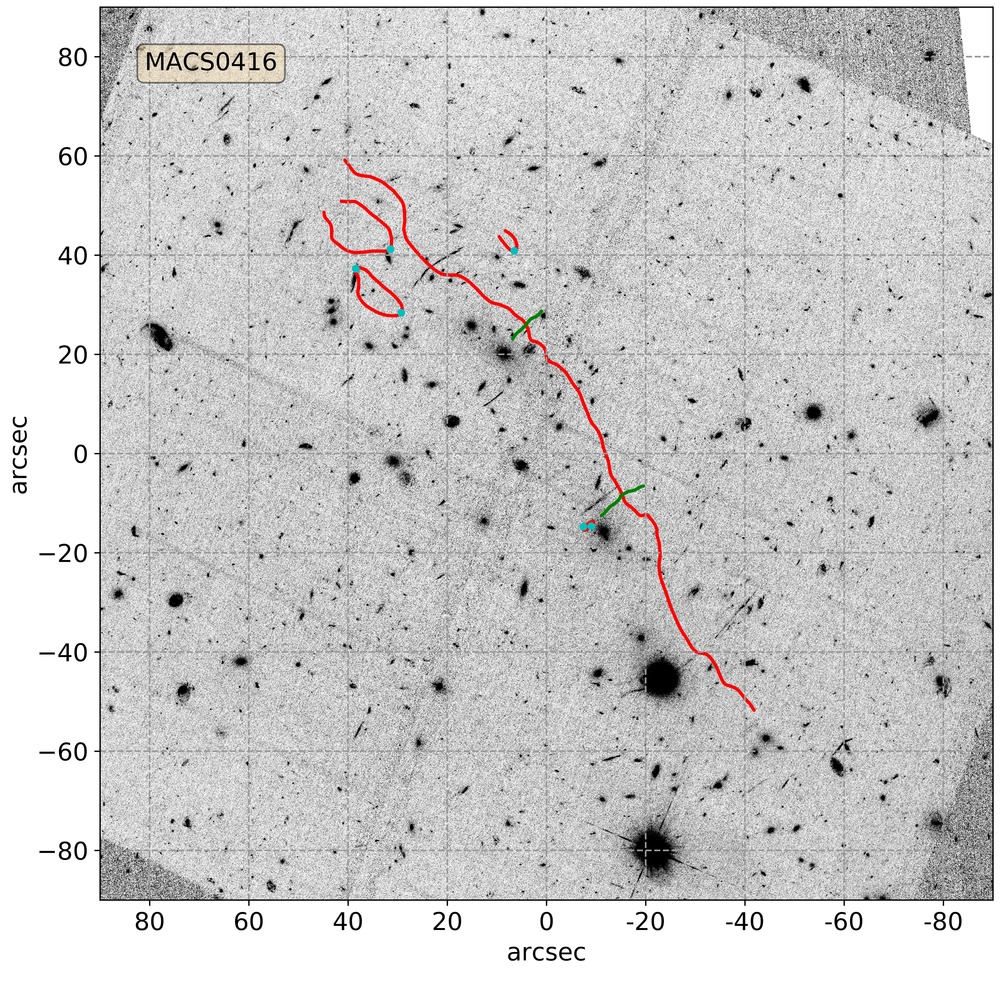}
    \caption{Singularity maps for HFF clusters:
  Every panel represents the singularity map for one of the HFF clusters. The name
  of the corresponding cluster is written in the upper left corner. In each panel, the red
  and green lines represent the $A_3$-lines corresponding to the $\alpha$ and $\beta$
  eigenvalues of the deformation tensor. The blue points represent the location of
  (hyperbolic and elliptic) umbilics. The cyan and magenta points represent the
  swallowtail singularities corresponding to the $\alpha$ and $\beta$ eigenvalues 
  of the deformation tensor. In each panel, the background is the cluster image in
  F435W band.}
  \label{fig: HFF singularity}
\end{figure*}
\begin{figure*}
  \ContinuedFloat 
  \includegraphics[width=\textwidth,height=8.0cm,width=8.0cm]{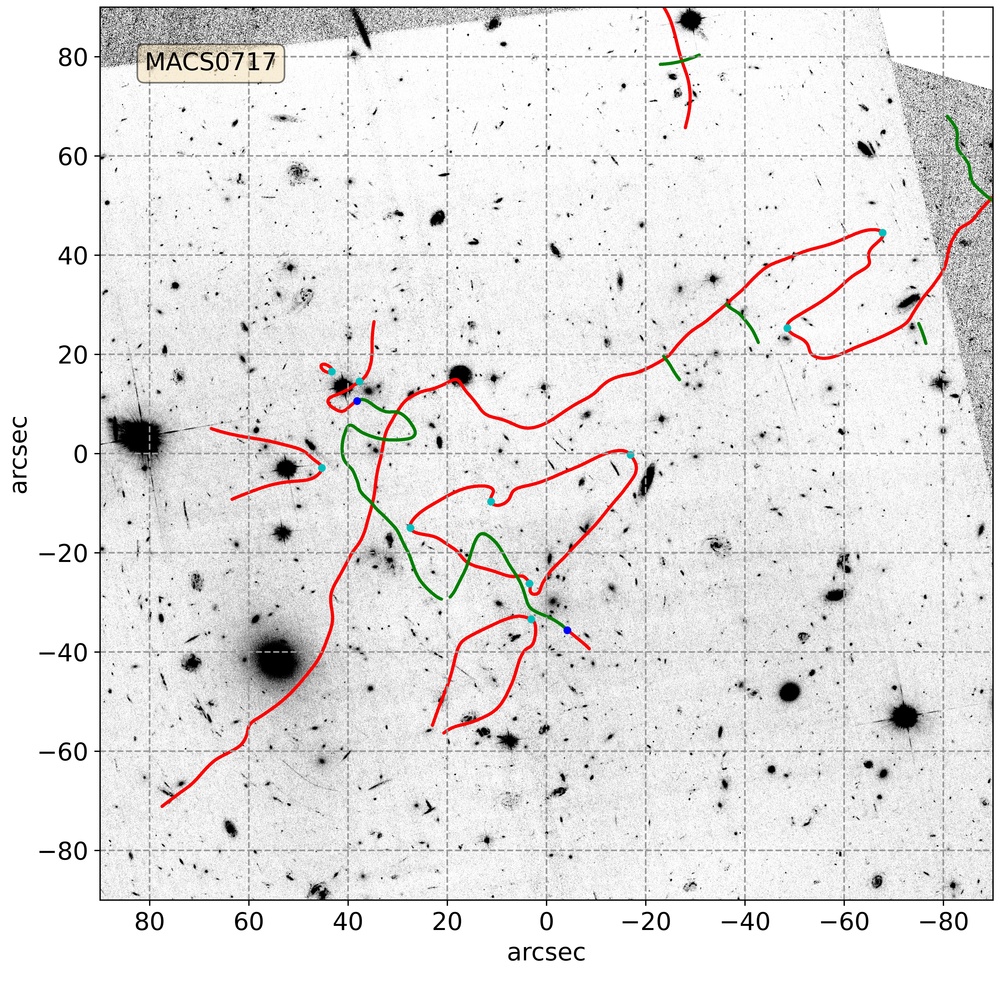}
  \includegraphics[width=\textwidth,height=8.0cm,width=8.0cm]{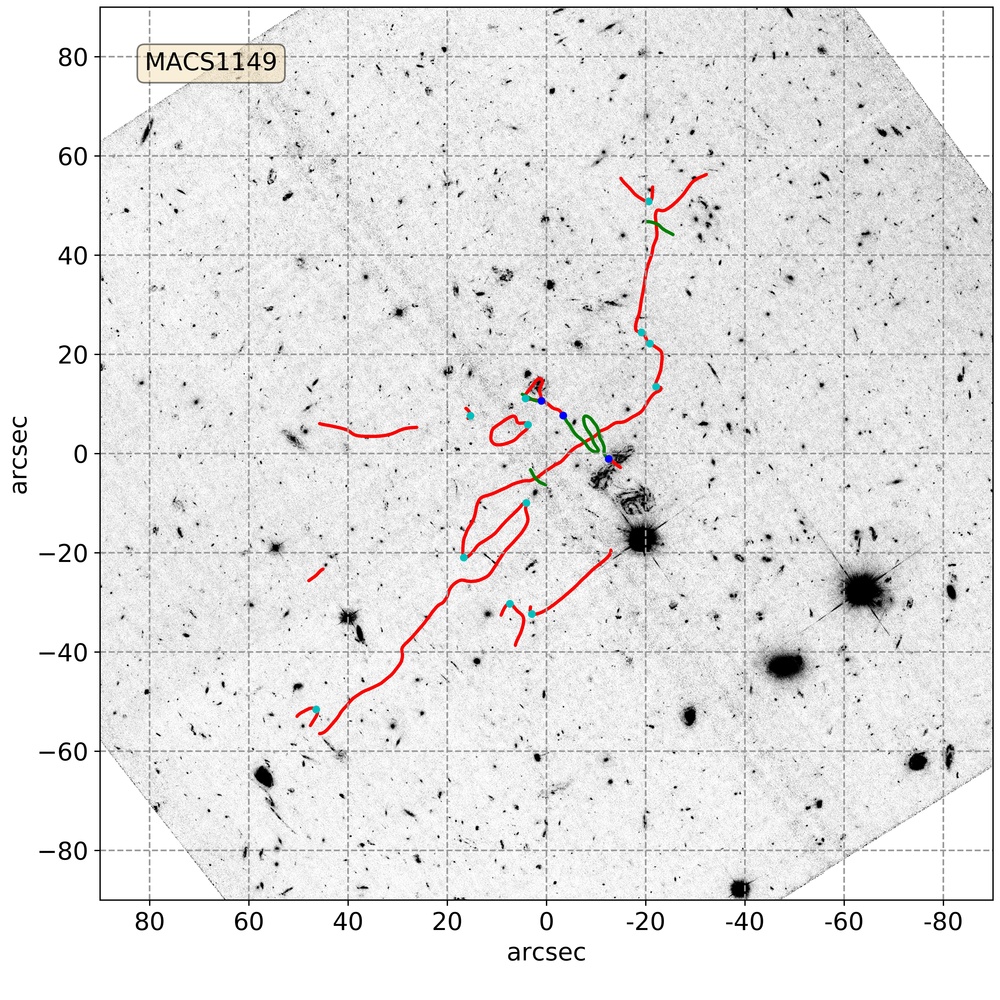}
  \caption{Cont.}
\end{figure*}

\begin{figure*}
  \includegraphics[width=\textwidth,height=7.0cm,width=8.0cm]
  {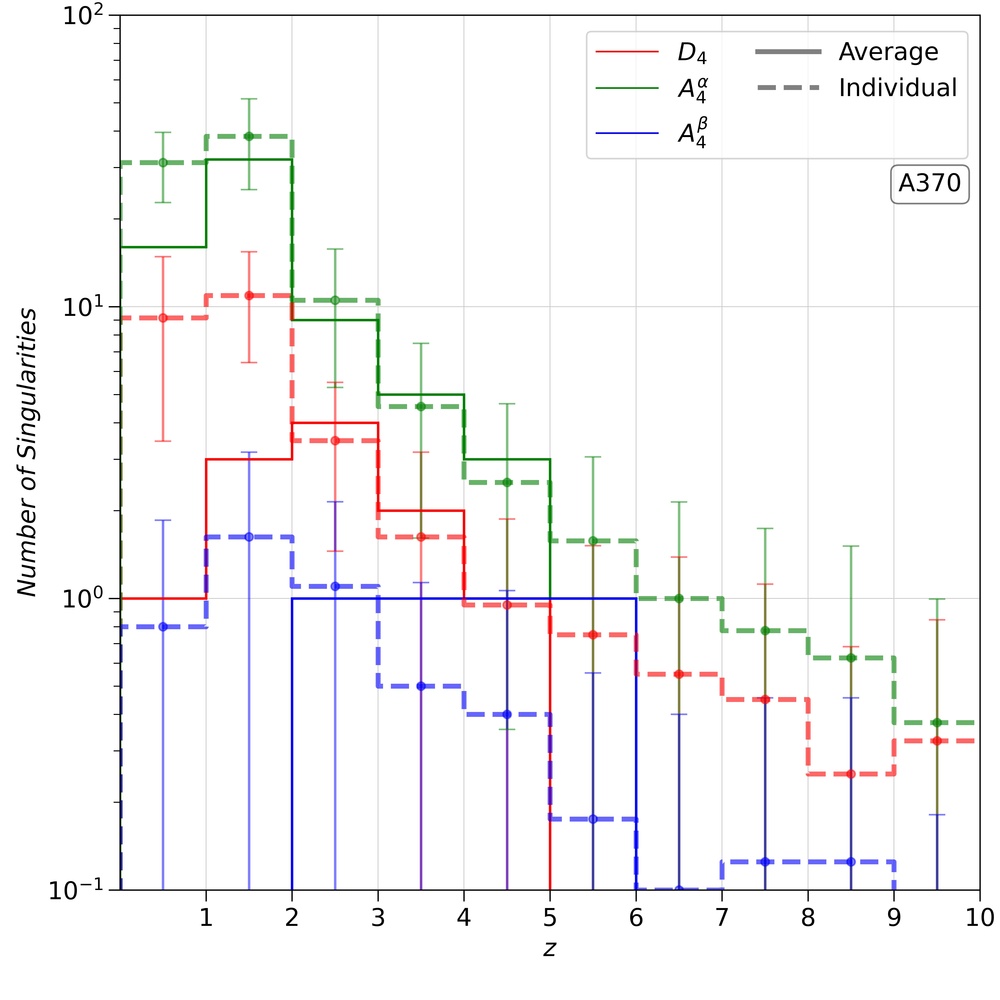}
  \includegraphics[width=\textwidth,height=7.0cm,width=8.0cm]
  {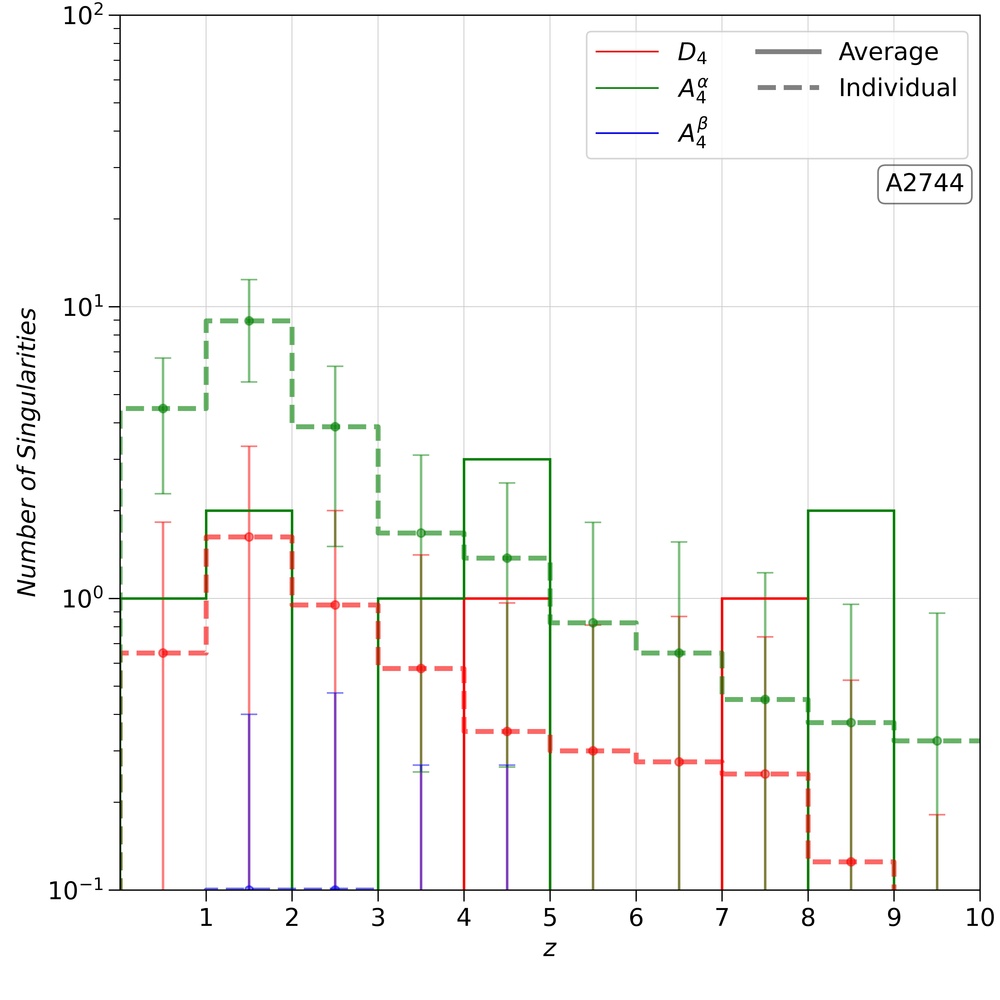}
  \includegraphics[width=\textwidth,height=7.0cm,width=8.0cm]
  {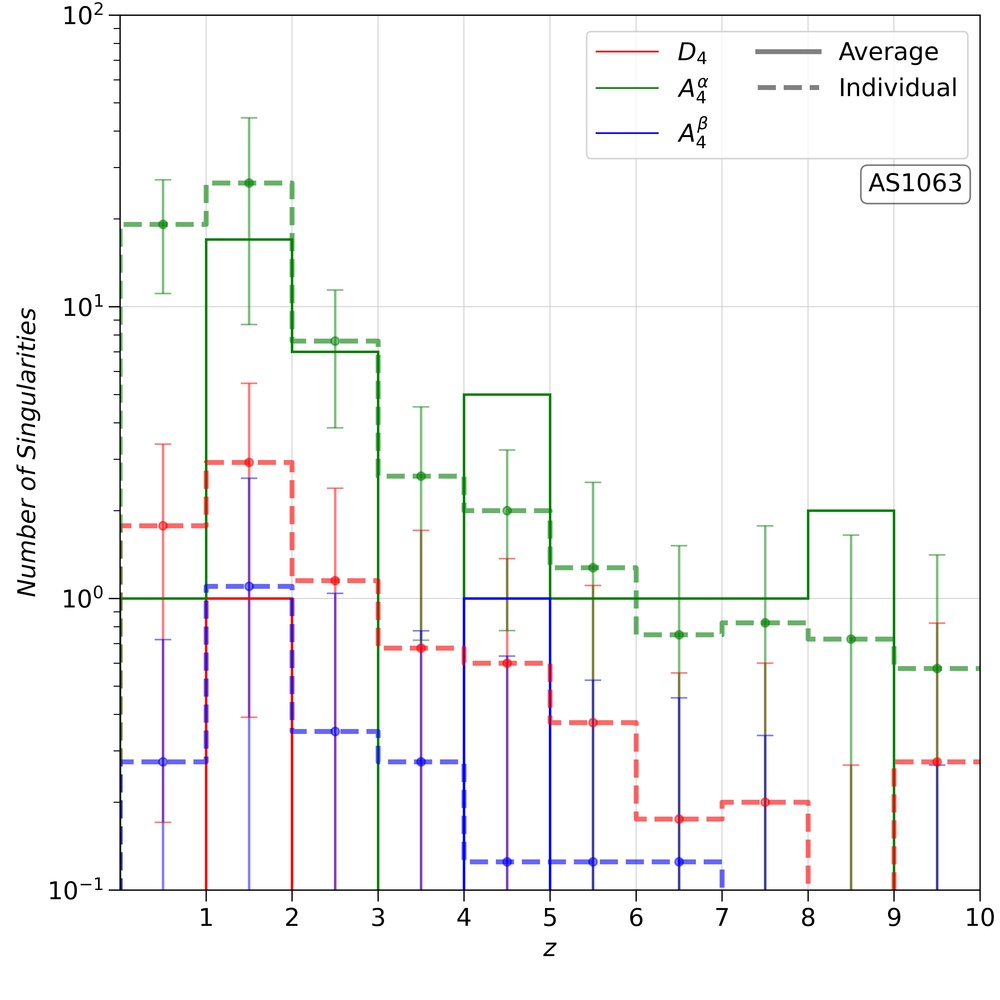}
  \includegraphics[width=\textwidth,height=7.0cm,width=8.0cm]
  {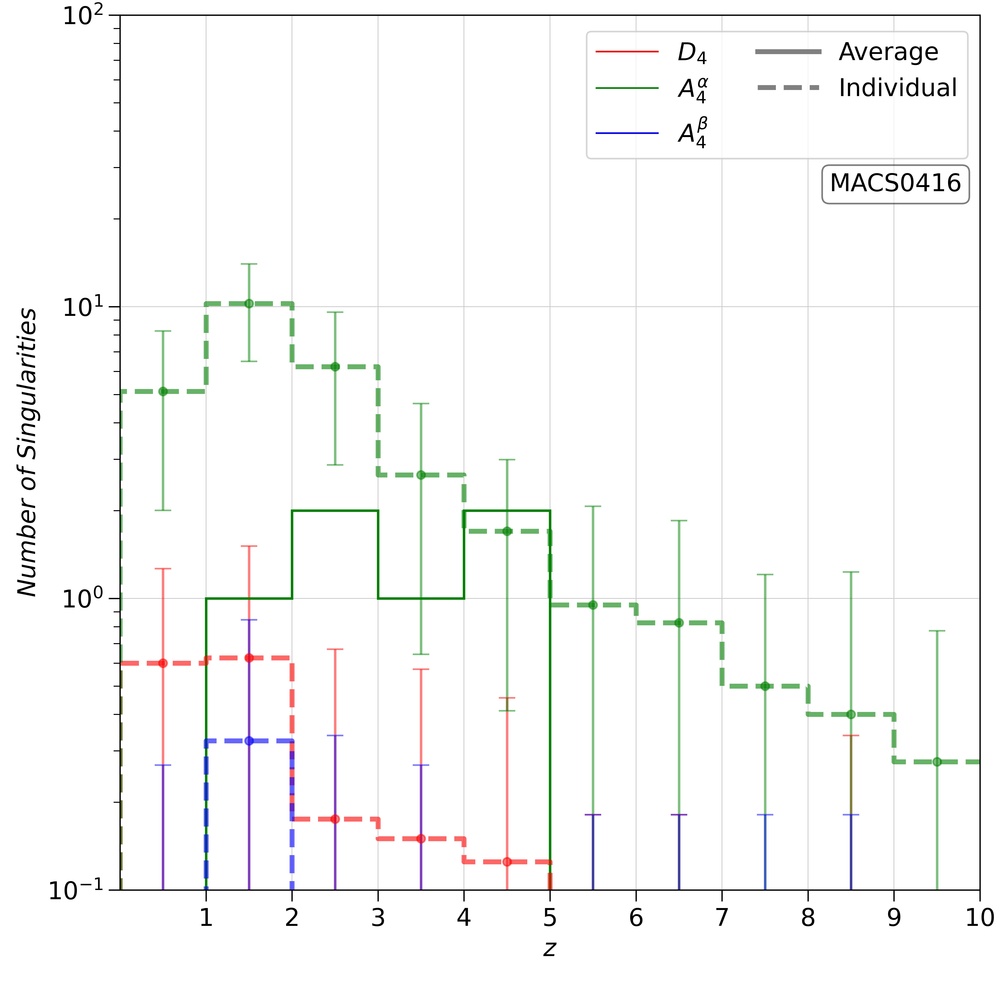}
  \includegraphics[width=\textwidth,height=7.0cm,width=8.0cm]
  {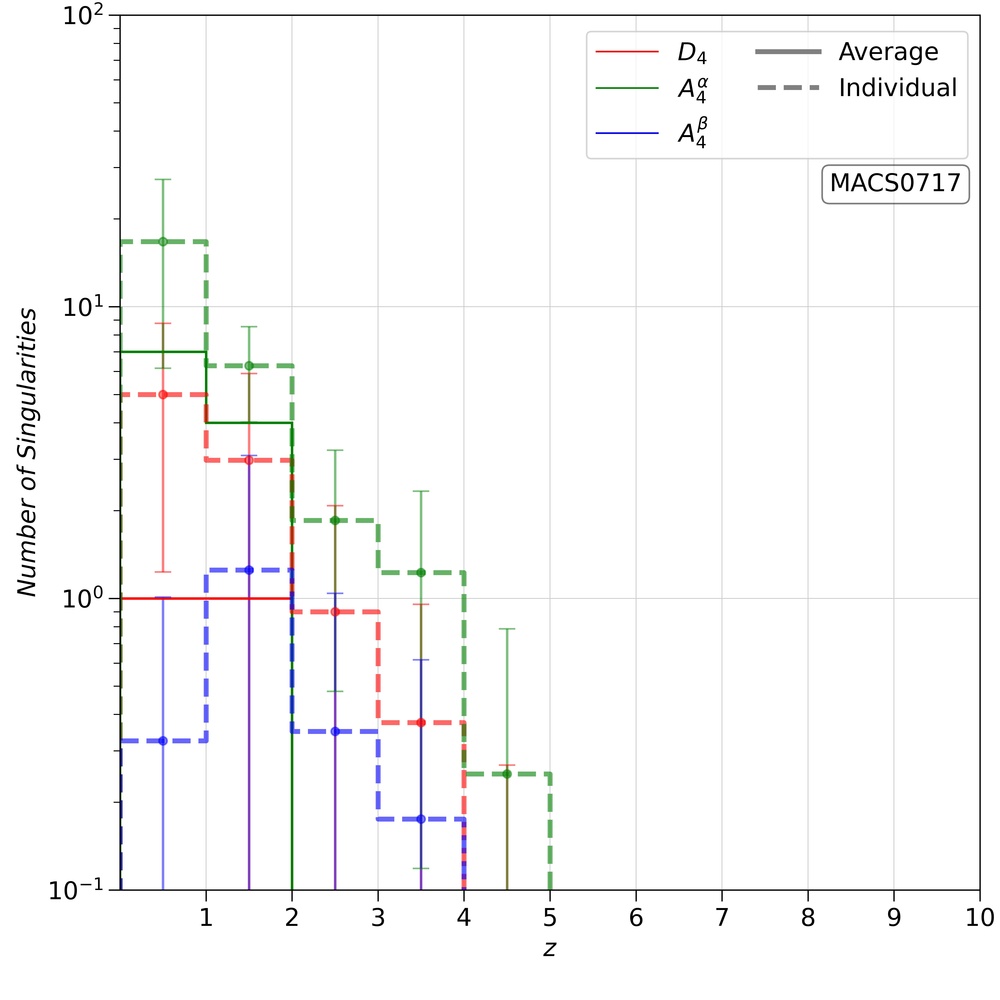}
  \includegraphics[width=\textwidth,height=7.0cm,width=8.0cm]
  {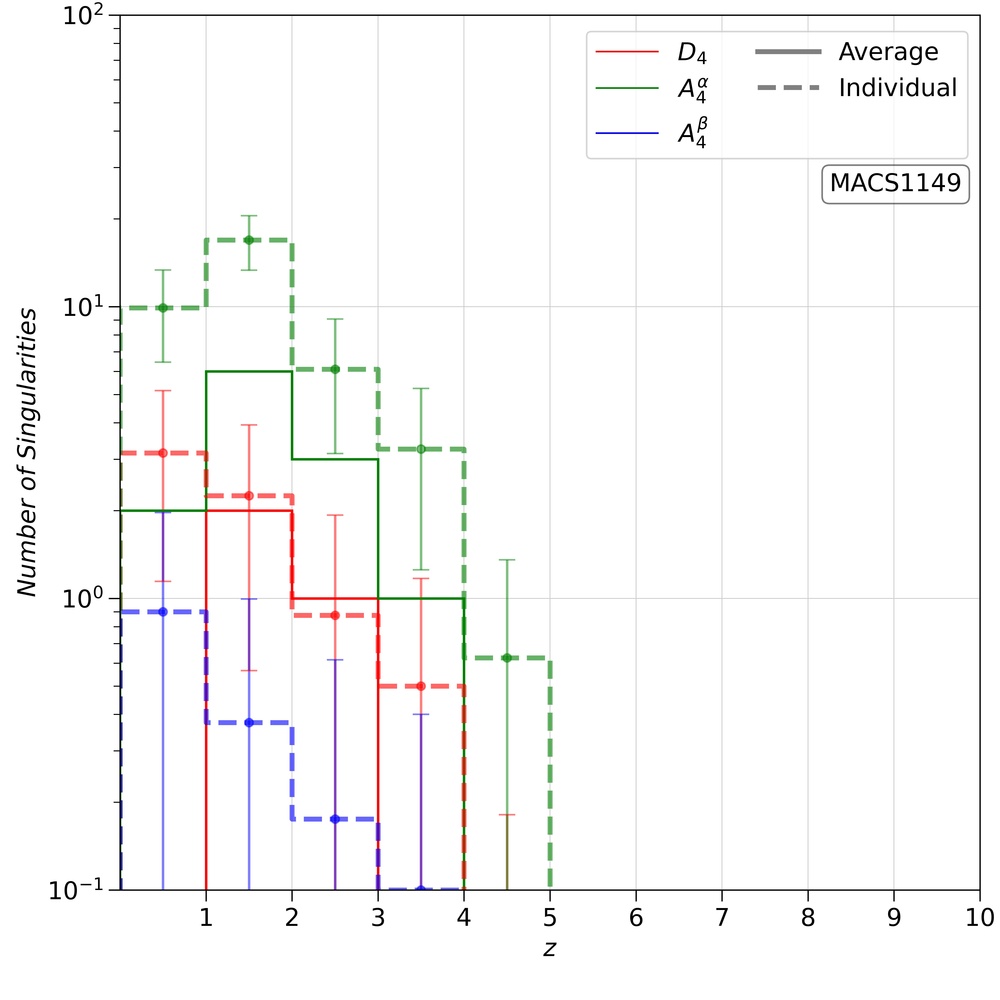}
  \caption{Number of singularities as a function of redshift in the HFF clusters:
  In each panel, the thin lines represent the number of singularities corresponding
  to the best-fit average mass model for the HFF clusters. 
  The thick dashed lines show the average number of singularity in individual mass
  models for the HFF clusters. The error bars represent the corresponding one sigma 
  scatter. The red lines represent the distribution of (hyperbolic$+$elliptic) umbilics.
  The green and blue lines represent the distribution of swallowtail singularities
  corresponding to $\alpha$ and $\beta$ eigenvalues of the deformation tensor.}
  \label{fig: HFF histogram}
\end{figure*}

\begin{figure*}
  \includegraphics[width=\textwidth,height=7.0cm,width=8.0cm]
  {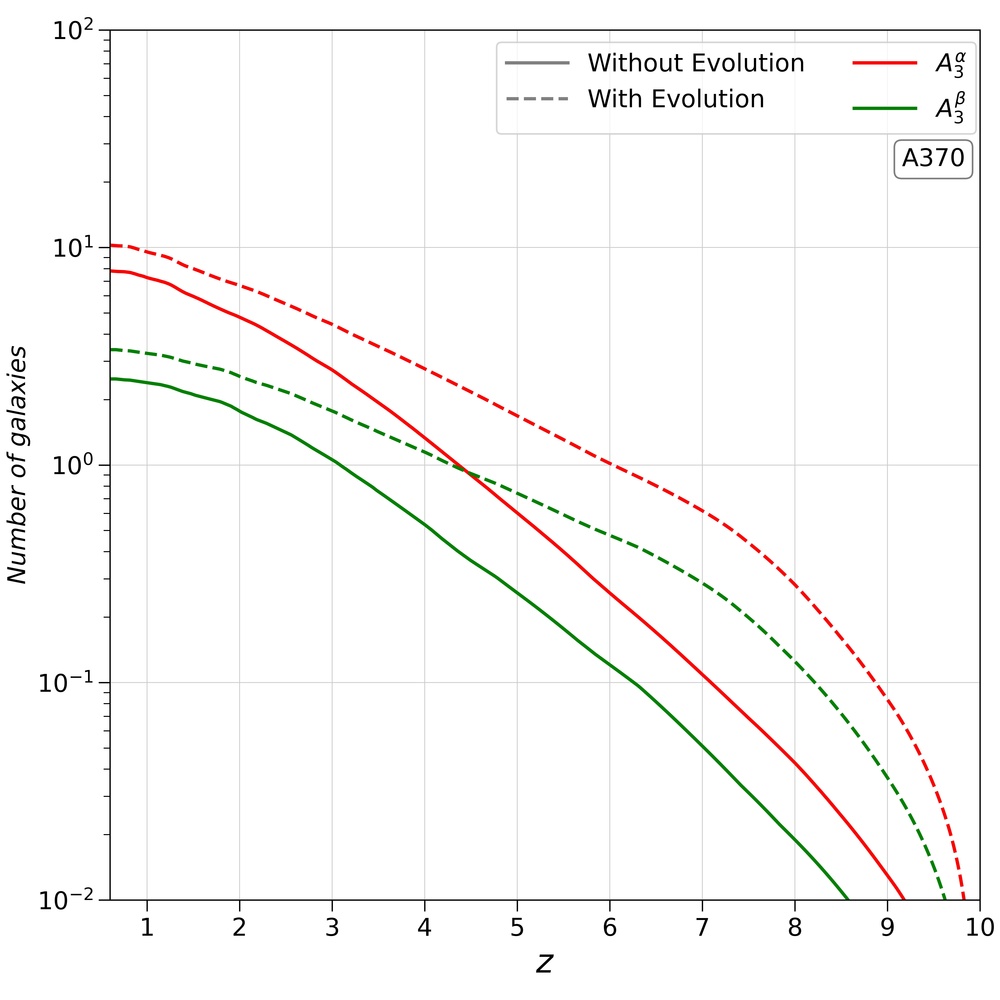}
  \includegraphics[width=\textwidth,height=7.0cm,width=8.0cm]
  {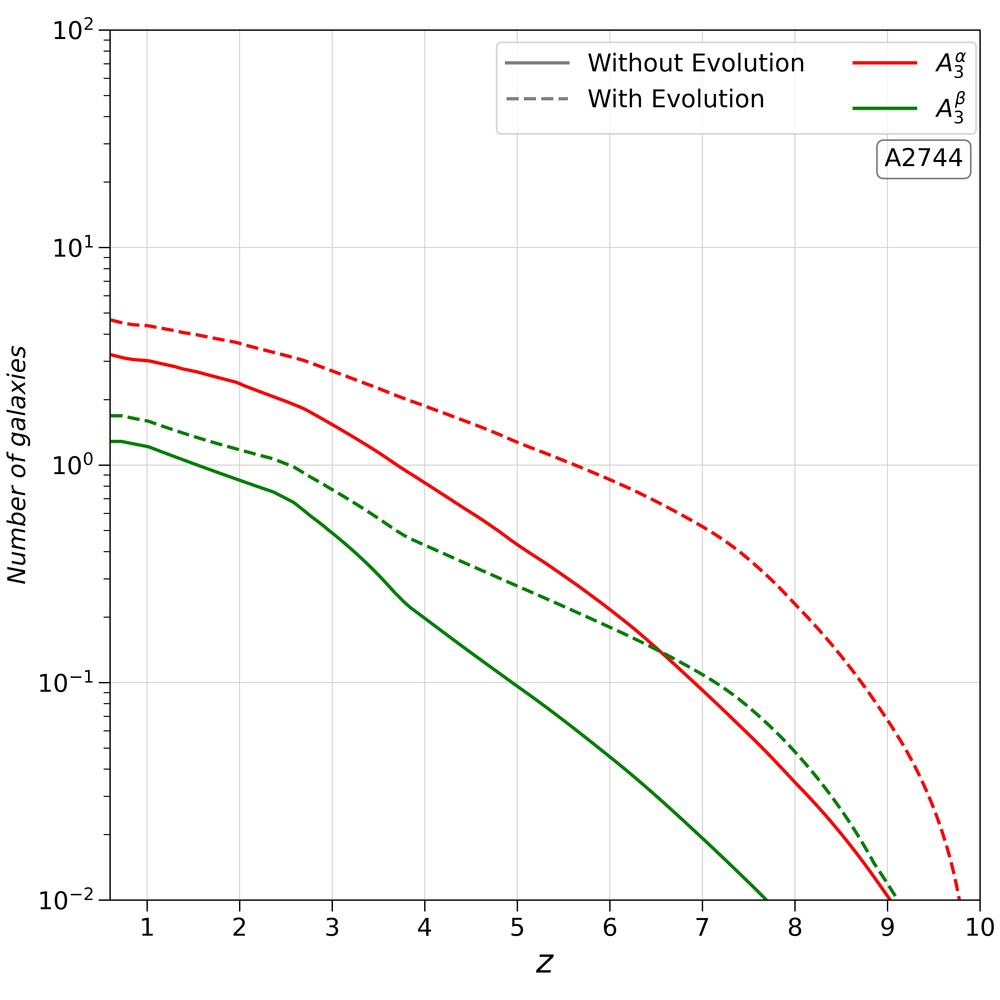}
  \includegraphics[width=\textwidth,height=7.0cm,width=8.0cm]
  {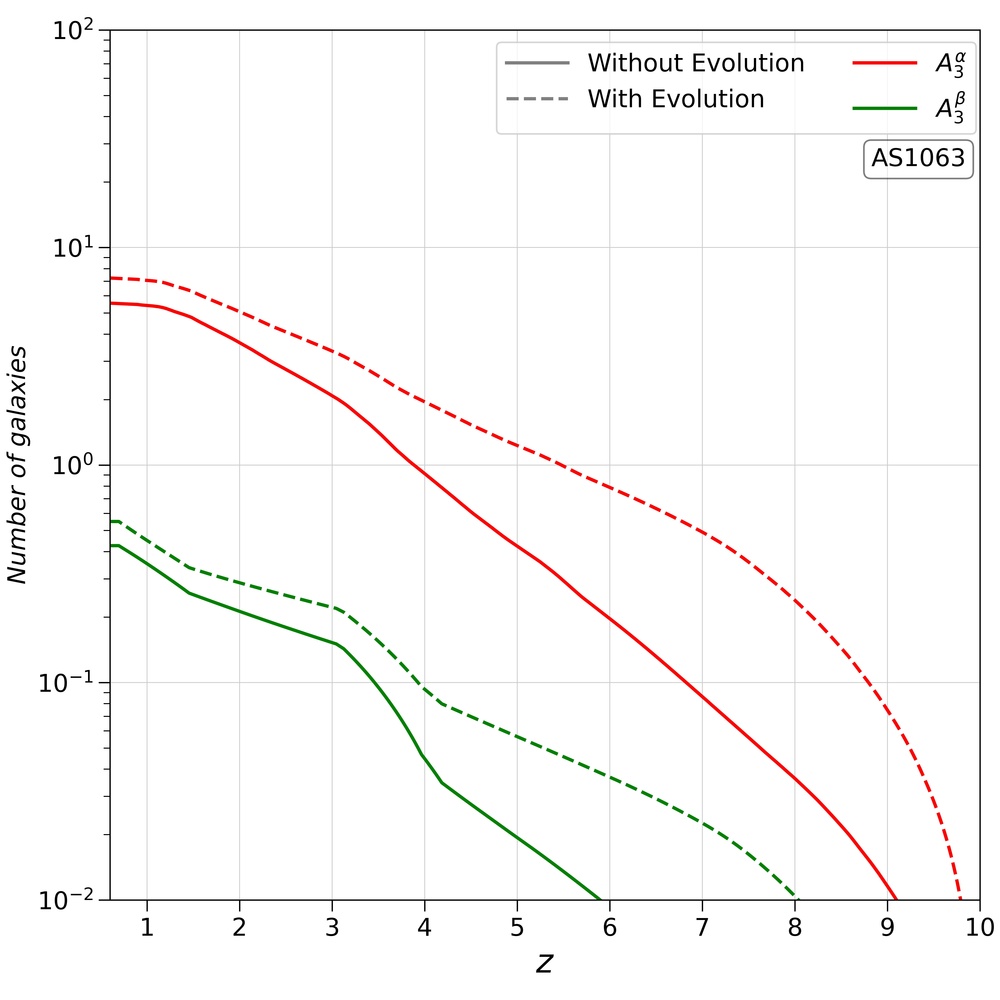}
  \includegraphics[width=\textwidth,height=7.0cm,width=8.0cm]
  {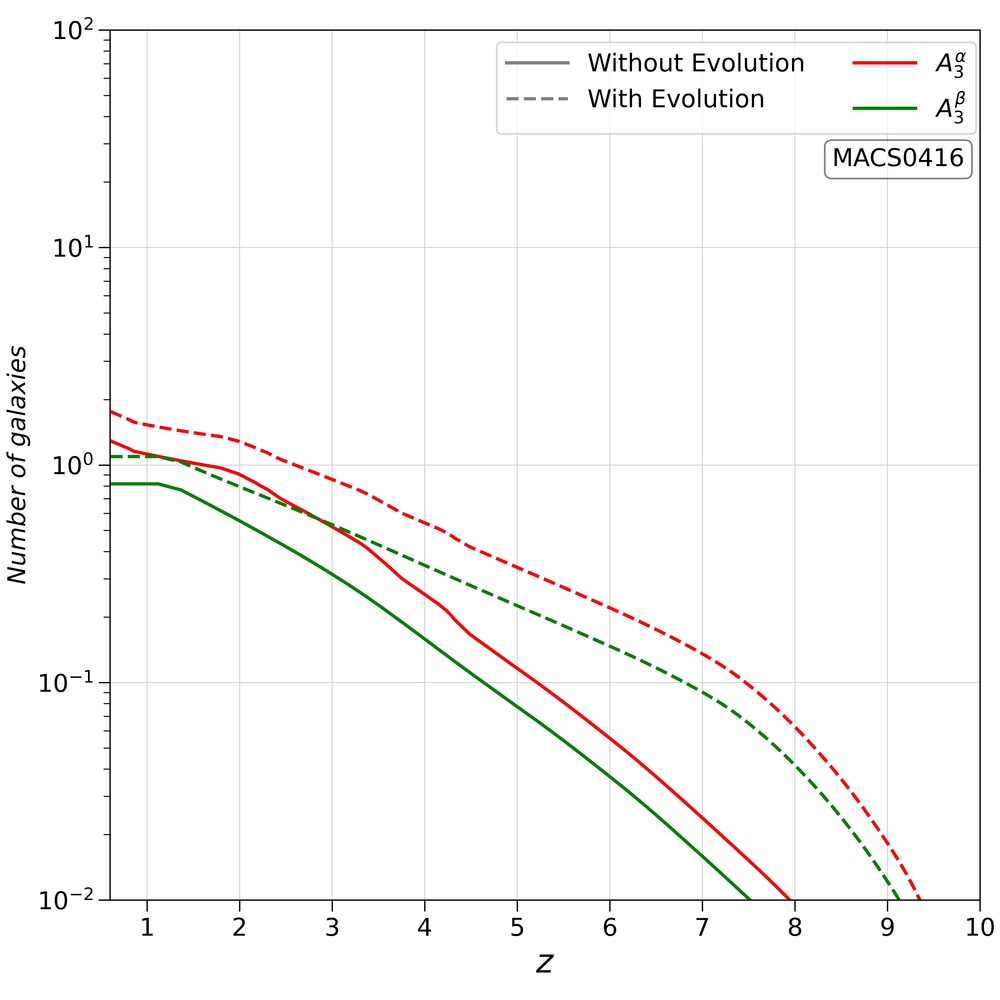}
  \includegraphics[width=\textwidth,height=7.0cm,width=8.0cm]
  {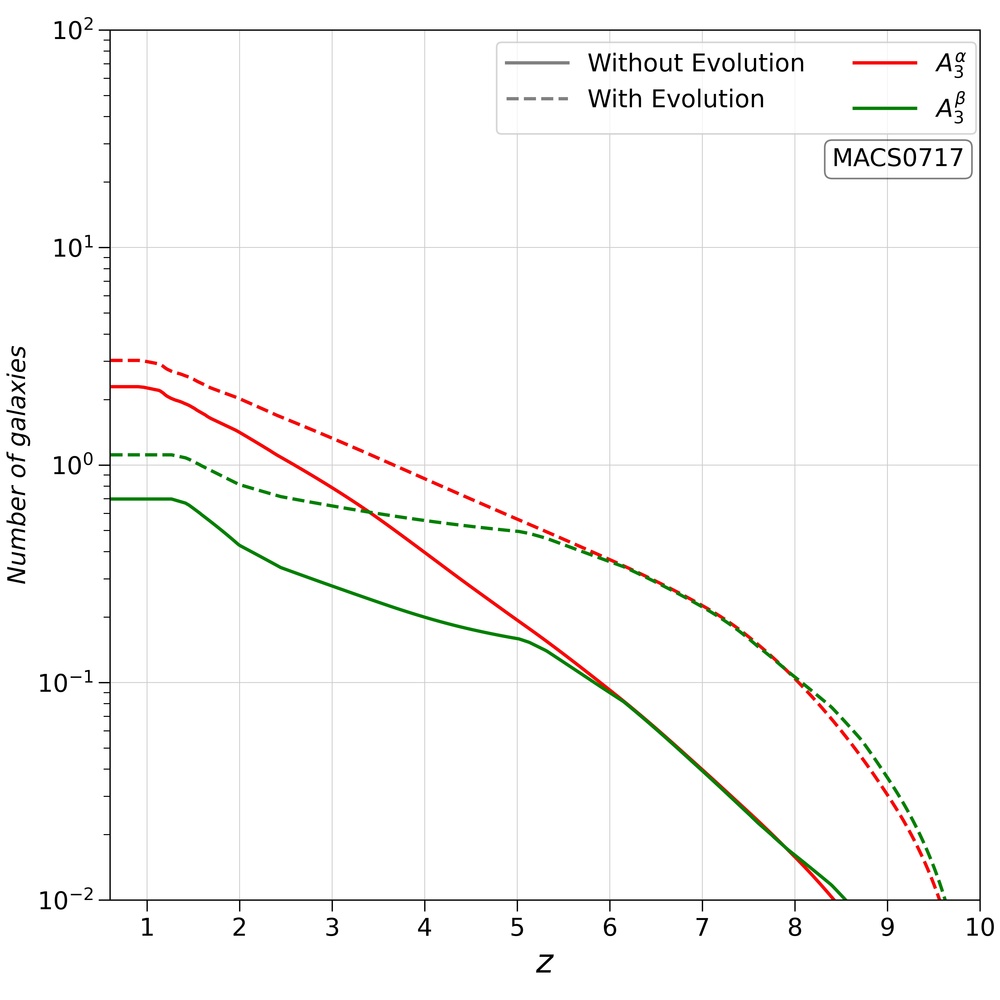}
  \includegraphics[width=\textwidth,height=7.0cm,width=8.0cm]
  {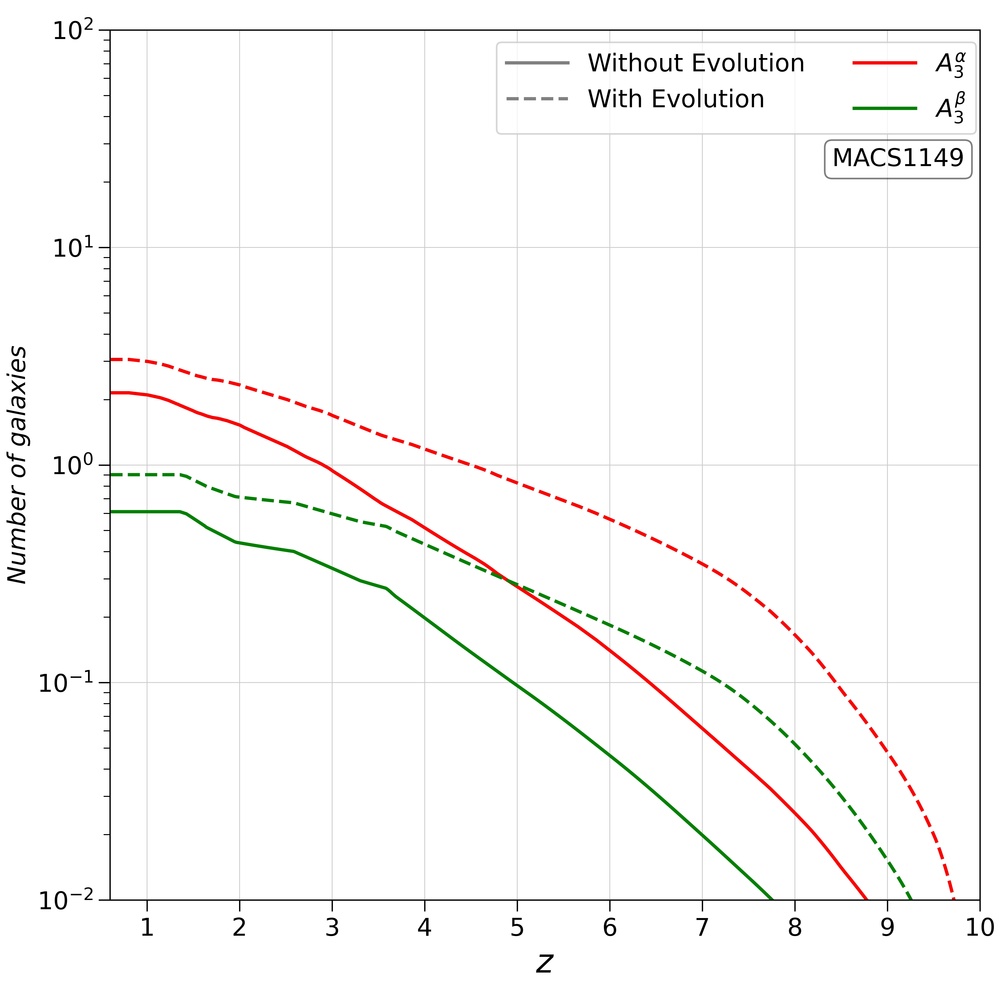}
  \caption{The cumulative number of source galaxies near the (tangential and radial)
  cusps as a function of redshift for HFF cluster lenses: The $y$-axis shows 
  the number at redshifts higher than $z$. Different panels are corresponding 
  to different singularity maps in Figure~\ref{fig: HFF singularity}, respectively. 
  The solid lines represent the galaxy numbers calculated using the fiducial model 
  used in C18, whereas the dashed lines indicate the galaxy numbers calculated 
  using the model with evolving feedback (please see C18 for more details). 
  The red and green lines denote the cumulative galaxy numbers corresponding to the
  tangential and radial cusps (corresponding to $\alpha$ and $\beta$
   eigenvalues of the deformation tensor).}
  \label{fig: HFF arc cross}
\end{figure*}

\begin{figure}
  \includegraphics[width=\textwidth,height=7.0cm,width=8.5cm]
  {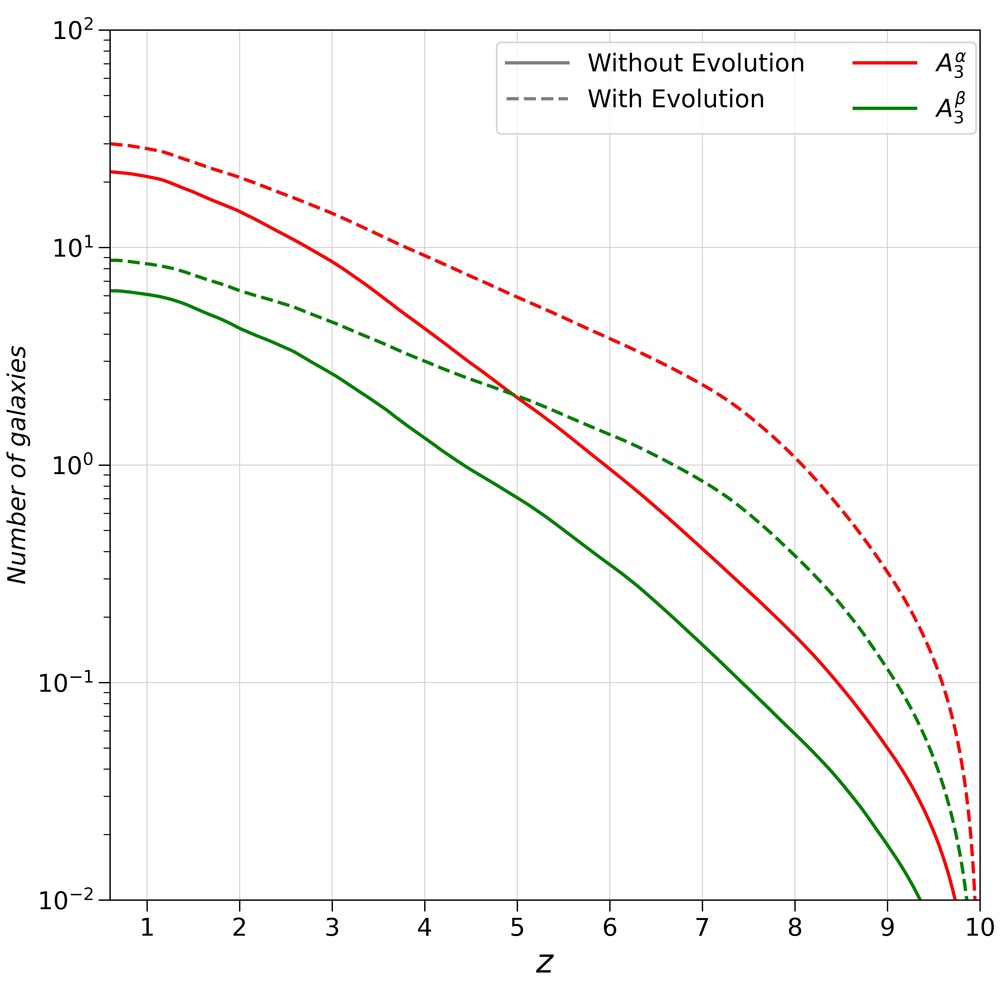}
  \caption{The total cumulative number of the source galaxies near tangential
  and radial cusps as a function of redshift for the HFF clusters: The $y$-axis 
  shows the number at redshifts higher than $z$. 
   Similar to Figure~\ref{fig: HFF arc cross}, the solid lines represent the 
   galaxy numbers calculated using the fiducial model used in C18, whereas the 
   dashed lines indicate the galaxy numbers calculated using the model with 
   evolving feedback. 
   The red and green lines denote the cumulative galaxy numbers corresponding 
   to the tangential and radial cusps (corresponding to $\alpha$ and $\beta$
   eigenvalues of the deformation tensor).}
  \label{fig: HFF arc cum}
\end{figure}

In MB21, it has been shown that the parametric and non-parametric 
mass models for a given HFF cluster show a significant difference 
in both cusp and point singularity cross-section. 
However, each mass model has uncertainties associated with it due 
to the finite amount of observational data.
As a result, these uncertainties also affect the results 
derived from the reconstructed mass models.
In this section, we study the effect of these uncertainties on the 
singularity maps corresponding to reconstructed mass models for 
simulated galaxy clusters.
We have considered two simulated clusters, Irtysh I and II 
from GWL20.
GWL20 reconstructed free-form mass models using \textsc{grale} for both Irtysh I 
and II with three different sets of multiple images, 150, 500, 1000.
The lens mass reconstruction with 150 images corresponds to the 
current observational scenario, whereas the 500 and 1000 multiple image 
cases correspond to future observations with the \textit{Hubble 
Space Telescope} (HST) and \textit{James Webb Space Telescope} (JWST).
For each case, individual mass models are obtained from forty different 
and independent \textsc{grale} runs and the final best-fit mass model is an 
average of the mass distributions from these forty independent \textsc{grale} runs. 
While constrained by the required computational resources this number is 
consistent with previous works with \textsc{grale} and it sufficiently spans 
the range of uncertainties in the reconstructed mass models. 
We refer the reader to look into GWL20 for more details.
Hereafter, for simplicity, the average mass models for 1000, 500, 150 
image cases for Irtysh I/II will be referred as Irtysh IA/IIA, IB/IIB, IC/IIC.

\subsection{Simulated Clusters: Singularity Maps}
\label{ssec: simulated singularity maps}

Once we have all the reconstructed mass models, we construct the 
singularity maps for all individual mass models and also for the final 
averaged one and the original mass models.
To construct the singularity maps, we chose a resolution of $0.06''$
in the lens plane to calculate relevant quantities.
As discussed in MB21, for mass models reconstructed using \textsc{grale},
such a resolution of mass maps is adequate for constructing singularity
maps.
The singularity maps cover sources upto a redshift of ten.
The singularity maps for the original and final averaged mass 
models for Irtysh I and II are shown in Figure~\ref{fig: irtysh I singularity} 
and~\ref{fig: irtysh II singularity}, respectively.
In each panel, the red and green lines represent the $A_3$-lines
corresponding to the $\alpha$ and $\beta$ eigenvalues of the deformation
tensor.
The (hyperbolic and elliptic) umbilics are shown by the blue points,
whereas the swallowtail singularities corresponding to $\alpha$ and 
$\beta$ eigenvalues are denoted by cyan and magenta points, respectively.
One example of singularity maps corresponding to individual runs for
Irtysh IA/IB/IC, IIA/IIB/IIC are shown in Figure~\ref{fig: Simulated singularity indi}.
For both Irtysh I and II, we see that final averaged mass maps are 
not able to recover the contribution of the marginally critical
structures in the singularity maps.
In a singularity map, such structures can be located by looking for the
isolated $A_3$-lines.
Even if some of the individual runs are able to recover contribution 
from such marginal structures, it may be possible that these structures
are not present in the final best-fit mass models due to the averaging over 
forty individual mass reconstruction.
On the other hand, it is also possible to have some additional contribution 
from the spurious structures which are not present in the original mass
distribution, as can be seen in Irtysh IIA in Figure~\ref{fig: irtysh II singularity}.

Apart from that, one can also notice differences between the $A_3$-line
structures in the singularity maps near the core region of Irtysh clusters
in the original and reconstructed mass models.
These differences may be a result of the fact that in the reconstruction
of Irtysh I and II, there are no  sources below redshift one. 
This difference is more significant between the original and individual
runs (Figure~\ref{fig: Simulated singularity indi}), but the averaging 
in the final best-fit mass model decreases this difference and brings the
reconstructed mass distributions closer to the original ones. 

\subsection{Simulated Clusters: Redshift Distribution of Singularities}
\label{ssec: simulated redshift distribution}

In this subsection, we study the redshift distribution of the point 
singularities for Irtysh I and II.
In Figure~\ref{fig: Irtysh histogram}, we represent the point singularities
for different mass models as a function of redshift.
Here we compare the number of singularities in the original Irtysh I 
and II mass models with the corresponding individual reconstructed 
mass models and with the corresponding final averaged mass model.
For example, the top-left panel of Figure~\ref{fig: Irtysh histogram}, shows
the number of singularities in the original Irtysh I (thin lines), 
the number of singularities in the final averaged Irtysh IA  mass 
model (thick dotted lines), and the average number of singularities 
in the corresponding individual runs (thick dashed lines).
The error bars represent the one-sigma scatter within these forty individual runs.
The red lines represent the (hyperbolic $+$ elliptic) umbilics.
Here, we are not discriminating between hyperbolic and elliptic
umbilics as the number of elliptic umbilics is very small compared to
the hyperbolic umbilics.
The green and blue lines represent the number of swallowtail 
singularities corresponding to $\alpha$ and $\beta$ eigenvalues
of the deformation tensor, respectively.

In each panel of Figure~\ref{fig: Irtysh histogram}, for source redshift $<2$,
we notice that the number of singularities in individual runs is 
significantly large compared to the number of singularities 
in the original mass Irtysh mass models, which is also evident from the 
individual runs in Figure~\ref{fig: Simulated singularity indi}.
This is due to the fact that \textsc{grale} introduces a significant number 
of structure (which decrease as the number of lensed images increase) 
at different positions in each individual
reconstruction, and as GWL20 do not have any sources at redshift $<1$, 
this effect is more significant in the central regions.
As we move towards the higher source redshifts, the difference between
original and individual run starts to narrow down. 
Although, the individual runs still predict slight excess of point
singularities compared to the actual mass models.

Although the individual runs give a significant excess of point
singularities below source redshift 2, such excess does not
occur in the final averaged mass models.
This behavior is expected from the fact that the averaging over
multiple individual runs smooths out spurious structures and
bring the final reconstructed mass model closer to the original
mass model, although most of the time with slight underestimation
of the point singularities.
The averaging over multiple realizations sometimes also introduces 
a small number of spurious point singularities at high redshifts, 
for example, as seen in the Irtysh IA, Irtysh IIB, and Irtysh IIC 
panels in Figure~\ref{fig: Irtysh histogram}.
However, the number of these spurious singularities is not significant
($\lesssim 2$ from Figure~\ref{fig: Irtysh histogram}).

\section{HFF Clusters}
\label{sec: hff}

Under the \textit{Hubble Frontier Fields (HFF) 
Survey}\footnote{\url{https://archive.stsci.edu/prepds/frontier/}}
program, \textit{Hubble Space Telescope} observed a total of six 
massive merging clusters (see~\citealt{2017ApJ...837...97L}
for more details).
For every HFF cluster, different groups have reconstructed mass
models using parametric, non-parametric and hybrid (a combination of
parametric and non-parametric) methods~\citep[e.g.,][]{
2007MNRAS.375..958D, 2005A&A...437...39B, 
2015MNRAS.452.1437J, 2014ApJ...797...48J, 2014ApJ...797...98L, 
2011MNRAS.417..333M, 2016MNRAS.459.1698M, 2010PASJ...62.1017O, 
2014MNRAS.437.2642S, 2018MNRAS.480.3140W, 2013ApJ...762L..30Z}.
To study the statistical uncertainties on the singularity maps, we
consider the non-parametric HFF cluster mass models.
These mass models for HFF clusters are reconstructed using the 
free-form method \textsc{grale}.
There are two reasons for choosing these mass models for the HFF clusters: 
\\
(1) The best-fit \textsc{grale} mass models for HFF clusters gives the simplest 
singularity maps compared to other techniques as shown in MB21 and 
puts the lower limit on the point singularity cross-section.
Hence, it is worthwhile to check how statistical uncertainties affect
the lower limit.
\\
(2) In MB21, we only considered the central $40''\times40''$ region
for the analysis. This choice was made due to the increasing number of
spurious singularities in the parametric mass models as we go away from
the center of the cluster.
Hence, it will not be useful to analyze the effect of uncertainties on
the singularity map in the presence of artifacts.
However, such a problem does not exist for these non-parametric mass models
as the singularity map is very simple and a resolution of $<0.1''$ is 
sufficient enough as shown in MB21.

Similar to the Irtysh clusters, for every HFF cluster, \textsc{grale} provides 
forty individual mass maps and one best-fit mass map that is the average of
these forty mass maps.
Depending on the number of images used in the reconstruction, there are
different versions of the reconstructed mass models available.
Here, we are using the v4 mass models for all of the HFF clusters,
and, for simplicity, we will be using the abbreviated names for the 
HFF cluster lenses.

\subsection{HFF Clusters: Singularity Maps}
\label{ssec: HFF singularity maps}

Following the procedure used for simulated clusters, we reconstruct 
the singularity maps for individual and the best-fit mass models of the 
HFF clusters.
The best-fit mass model singularity maps for the HFF clusters are shown in 
Figure~\ref{fig: HFF singularity}.
Here again, the singularity maps are drawn for sources upto redshift ten.
The red and green lines show the $A_3$-lines corresponding to the $\alpha$
and $\beta$ eigenvalues of the deformation tensor.
The blue, cyan, and magenta points represent the (hyperbolic and elliptic) 
umbilics, swallowtail for $\alpha$, and swallowtail for $\beta$ eigenvalue, 
respectively. 
The online available v4 mass models have a resolution of $\geq \: 0.2''$.
However, as the \textsc{grale} mass models are the superposition of a large 
number of projected Plummer density profiles, one can (in principle) resolve 
them up to any arbitrary resolution.
In our current work, we use a resolution of $0.06''$ for all of the HFF clusters. 
Although, as mentioned above, a resolution of $\sim0.1''$ is sufficient
for \textsc{grale} mass models but, in order to be more certain that we did not
miss any point singularities, we chose a resolution of $0.06''$ (please
see discussion about stability of point singularities in MB21).

We can see in Figure~\ref{fig: HFF singularity} that the A370 yields a very 
complex singularity map compared to other HFF clusters.
Hence, the cross-section for image formation near a cusp or a point 
singularity is maximum for the A370 in the HFF clusters.
One can also see that the point singularity cross-section 
calculation done in MB21 was based on the central region of 
$40''\times 40''$; however, the full singularity map extends upto a much
bigger region in the lens plane for a source redshift of ten.
Hence, the estimations done in MB21 are actually underestimates.
Due to consideration of the central region, the cross-section of the 
hyperbolic umbilics was more than the swallowtails corresponding to the
$\alpha$ eigenvalues. 
However, looking at the complete singularity maps in 
Figure~\ref{fig: HFF singularity}, one can see that the cross-section for
the image formation near swallowtails corresponding to $\alpha$ eigenvalue
is higher compared to the hyperbolic umbilics as the outer regions of the
singularity maps show more swallowtails than umbilics.  
This also explains why we observed a larger number of images near swallowtails
(corresponding to $\alpha$ eigenvalue) than hyperbolic umbilics. 
Here, we do not repeat that calculation and consider the earlier estimates
as a lower limit.

Comparing the singularity maps of HFF clusters with each other
one can also see that not all cluster lenses contribute equally
in the point singularity or arc cross-section.
For example, we can see that A370 is roughly five times more
efficient in producing image formation near a swallowtail or a cusp
compared to MACS0416.
The efficiency of other HFF clusters lies between MACS0416 and A370.
Such a difference is also observed in the corresponding magnification 
maps~\citep[e.g.,][]{2014ApJ...797...48J, 2019MNRAS.486.5414V}. 
However, one should keep in mind that the area in the source plane magnified 
by a factor $\mu\geq\mu_{th}$ has a significant contribution from the fold caustic. 
Hence, having a larger source plane area magnified by $\mu\geq\mu_{th}$ 
does not directly imply that the corresponding point singularity cross-section 
will also be higher.
One can still use the magnification maps to get a qualitative idea about the arc 
and point singularity cross-section since large critical lines in the lens plane 
mean having more probability of a substructure introducing distortions in it and 
introducing extra cusps in the source plane.
We would like to remind the reader that these inferences are based on 
specific models of HFF clusters, and these numbers can vary based on 
the reconstruction method used.

\subsection{HFF Clusters: Redshift Distribution of Singularities}
\label{ssec: HFF redshift distribution}

Unlike simulated clusters, we do not have the actual mass
distribution of a real galaxy cluster.
Hence, for HFF clusters, we compare the point singularities in the 
individual runs with the point singularities in the corresponding
final best-fit mass models.
The redshift distribution of point singularities in the HFF clusters
is shown in Figure~\ref{fig: HFF histogram}.
In each panel, the red, green, and blue lines represent the number
of (hyperbolic $+$ elliptic) umbilics, the number of swallowtail
singularity for $\alpha$, and the number of swallowtail
singularity for $\beta$ eigenvalues, respectively.
Here, we do not differentiate between hyperbolic and elliptic umbilics
as the number of elliptic umbilics is negligible as compared to the 
hyperbolic umbilics.
In each panel, the thin solid lines represent the number of point 
singularities corresponding to the averaged best-fit mass models,
and the thick dotted lines represent the average number of point
singularities in the individual runs.
The errorbars associated with thick dotted lines represent the 
one-sigma scatter in the number of point singularities in the 40
realizations.

As with the simulated Irtysh clusters, the average number of
point singularities in the individual runs is higher
(more than one sigma difference in some clusters) compared to the best-fit 
mass models at redshifts $z< 2$.
In Irtysh clusters, we expected to observe such discrepancy in the central
regions of the clusters as there were no sources below redshift one
that can provide lensed images near the central region in order
to constrain it better.
For A370 and AS1063, there are lensed images below redshift
one, but we still observe an excess of point singularities in the
individual runs.
Hence, we can say that introducing additional images below redshift
one may help in decreasing the discrepancy in the central images.
But we cannot be sure that it will always bring the numbers within
one sigma error bars.

As we go towards higher source redshifts, the number of point
singularities decreases in individual runs.
However, the same cannot be said for the best-fit mass models.
For example, in the case of A2744, there are some umbilics and swallowtails
that are present even at very high source redshifts.
Following the simulated Irtysh cluster, we may infer that these point
singularities at very high redshift are spurious features.
However, just by considering the \textsc{grale} mass models, we cannot be sure about 
that as there are no reference mass models for the real cluster lenses.
As the point singularities depend on the higher-order derivatives of the 
deformation tensor, taking inputs from the parametric reconstruction for 
validations of these features may not be very useful. 

\subsection{HFF Clusters: Arc Cross-Section}
\label{ssec: HFF arc cross section}

The $A_3$-lines constitute the backbone of a singularity map.
These lines trace the points in the lens plane that correspond to cusps 
in the source plane.
Hence, if we draw the critical curves for a given redshift on the top of the
$A_3$-lines in the lens plane then these critical curves cut the $A_3$-lines 
at the points that correspond to cusps in the source plane for that source redshift.
As a result, without looking into the source plane, one can count (in a very 
simple way) the number of cusps on tangential and radial caustics in the source 
plane.
A source lying near a cusp in the source plane leads to the formation of a 
three-image arc.
Hence, by using the $A_3$-lines and the critical curves, we can easily
calculate the three-image
(tangential and radial) arc cross-section for a given lens model
with a given population of sources.

To calculate the arc cross-section, we divided the source redshift range 
[0.6, 10] in equal intervals of $\triangle z=0.01$. 
After that, we draw the critical curves in the lens plane for each interval 
and calculate the number of points in the lens plane where a (tangential or 
radial) critical curve cuts the corresponding $A_3$-line. 
We assume that the number of such points is constant in the respective redshift 
interval of $\triangle z=0.01$. 
Similar to point singularities in MB21, we consider an area of 5 kpc in 
the source plane and calculate the number of source galaxies with the image 
formation near the cusp points. 
This method will underestimate the number of cusp points if two $A_3$-lines 
are very near to each other. 
Such a scenario occurs when a swallowtail singularity gets critical, and two 
cusps emerge in the source plane. 
However, the underestimation is not significant as the singularity maps for 
non-parametric models are not very complex.

Following this method, we calculated the cusp cross-section for HFF clusters. 
The galaxy source population is taken from C18 for one filter (F200W) of the JWST.
The results are shown in Figure~\ref{fig: HFF arc cross}.
The x-axis represents the source redshift, and the y-axis shows the number of 
source galaxies that are expected to give the image formation near cusps in the 
source plane higher than redshift $z$.
The red and green lines represent the number of source galaxies near the
tangential and radial arcs in the source plane. 
In each panel, the cross-section is plotted considering sources in the redshift
range [0.6, 10]. 
These cross-section plots again validate our inference from the singularity maps 
(in \S\ref{ssec: HFF singularity maps}) that A370 is most efficient in producing 
the tangential and radial arcs among the HFF clusters.
The cumulative arc cross-section for all clusters is shown in Figure~\ref{fig: HFF arc cum}.
From our analysis, we expect to observe at least 10-20 tangential and 5-10 radial
three image arcs with the JWST.
Again, as the \textsc{grale} mass models have a very small contribution from 
galaxy scale substructures in the cluster lenses, these values should be 
considered as a lower limit.
Since, \textsc{grale} provides a lower limit on the cusp cross-section, it will
not be very beneficial to add the one-sigma error bars corresponding to individual
runs in Figure~\ref{fig: HFF arc cross} and \ref{fig: HFF arc cum}.
The reason is that the length of $A_3$-lines in the individual runs is 
significantly larger than the averaged best-fit mass model and due to 
which these error bars will always be above the best fit lines.

\section{Conclusions}
\label{sec: conclusions}

In this work, we have studied the effect of statistical uncertainties
associated with cluster lens mass reconstruction with \textsc{grale} on 
point singularities.
We considered two simulated clusters, Irtysh I and II from GWL20 and
six galaxy clusters from the HFF survey.
For both simulated and real galaxy clusters, the mass models were
reconstructed using \textsc{grale}.
For each simulated galaxy cluster, we had three different mass models
reconstructed with different sets of multiple images (please see GWL20
for more details).
For each cluster lens, 40 reconstructions were done, and the best-fit
mass models were the average of them.
As a first step, we have constructed singularity maps for the original
mass models (in the case of Irtysh I and II), for all of the individual 
mass models and for the best-fit mass models.
In the next step, we have compared the singularity maps corresponding
to the individual realizations with the original and average best-fit
mass models.
These comparisons for simulated and HFF clusters are shown in Figure
\ref{fig: Irtysh histogram} and \ref{fig: HFF histogram}, respectively.

We find that the singularity maps corresponding to the best-fit mass models
have a relatively small number of point singularities compared to the
individual runs.
As a result, the best-fit mass models provide a lower bound on the
number of singularities that a lens has to offer.
For mass models reconstructed by \textsc{grale}, such a result is expected due
to the fact that individual realizations contain spurious features,
and the averaging to get the best-fit mass models is done in order
to remove the contribution from these spurious features.
Hence, the final best-fit mass model contains a low number of small scale
structure and low number of point singularities.
The differences in the number of point singularities between individual 
runs and the best-fit are maximum at source redshift $z<2$ for both simulated 
and real clusters.
One can expect such discrepancy to arise at these redshifts as the mass models
in the central regions of the clusters, at present, is not very well constrained.
However, in order to be sure about this explanation, one will need to 
construct the mass models for simulated clusters with multiple images near 
the cluster center.

Such an averaging (over 40 realizations) to get the final best-fit mass model
also simplifies the over $A_3$-line structures in the singularity maps.
As $A_3$-lines in the singularity maps correspond to the number of 
cusps in the source plane at every source redshift, the final best-fit mass model
also has a smaller number of cusps compared to individual realizations.
Hence, the final best-fit mass models also provides a lower limit on the
number of image formations near cusps.
In our current work, we have also calculated the cusp cross-section for the
HFF clusters using the best-fit mass models.
we find that A370 is most efficient in producing three image
arcs and the image formation near point singularities among the HFF 
clusters based on \textsc{grale} modeling.
In the HFF clusters, one would expect to observe at least 10-20 tangential
and 5-10 radial three image arcs with JWST.

The key take away from our current work is that the best-fit mass models
constructed using the \textsc{grale} provide the lower limit on the number of
cusps and point singularities for a given cluster lens. 
Folding in the statistical uncertainties does not decrease the numbers 
obtained from the best-fit mass models.
As shown in MB21, one would expect to observe at least one hyperbolic
umbilic and one swallowtail for every five clusters with JWST even when
we include the corresponding statistical uncertainties.
Apart from that, one also expects at least 10-20 tangential and 5-10 radial 
three image arcs with JWST in the HFF clusters.

The above results are based on the non-parametric modeling by \textsc{grale}, which
underestimates the contribution from the galaxy scale substructures in the
cluster lenses. 
Hence, if one considers parametric modeling, the above quoted numbers are
expected to increase.
As shown above, these singularity maps also provide a very nice way to compare 
the efficiency of different cluster lenses.
This feature can also be used to compare the simulated and real clusters
on qualitative (visual comparison of singularity maps) and quantitative 
basis (arc and point singularity cross-sections).

Apart from that, image formation corresponding to arcs or point singularities
shows three or more images lie in a very small region of the lens plane.
Hence, the time delay between these images is expected to be small compared
to other (typical) multiple image formation scenarios.
The other important feature of point singularities is their characteristic
image formation, which can be very helpful in locating them.
These possibilities are the subject of our ongoing work.

\section{Acknowledgements}

AKM would like to thank Council of Scientific $\&$ Industrial Research 
(CSIR) for financial support through research fellowship  No. 524007.
Authors thank the anonymous referee for useful comments.
This research has made use of NASA's Astrophysics Data System 
Bibliographic Service.
We acknowledge the HPC@IISERM, used for some of the computations 
presented here.
\section{Data Availability}

The simulated cluster (Irtysh I and II) data from GWL20 and the high 
resolution HFF cluster mass models corresponding to the Williams group
are available from the modelers upon reasonable request.


\appendix

\section{Individual Runs}
\label{sec: individual runs}

Figure \ref{fig: Simulated singularity indi}, shows one individual run for each
Irtysh IA, IB, IC, IIA, IIB, IIC.
Similarly \ref{fig: HFF singularity indi}, Shows six individual runs for A370.  

\begin{figure*}
  \includegraphics[width=\textwidth,height=7.4cm,width=7.5cm]{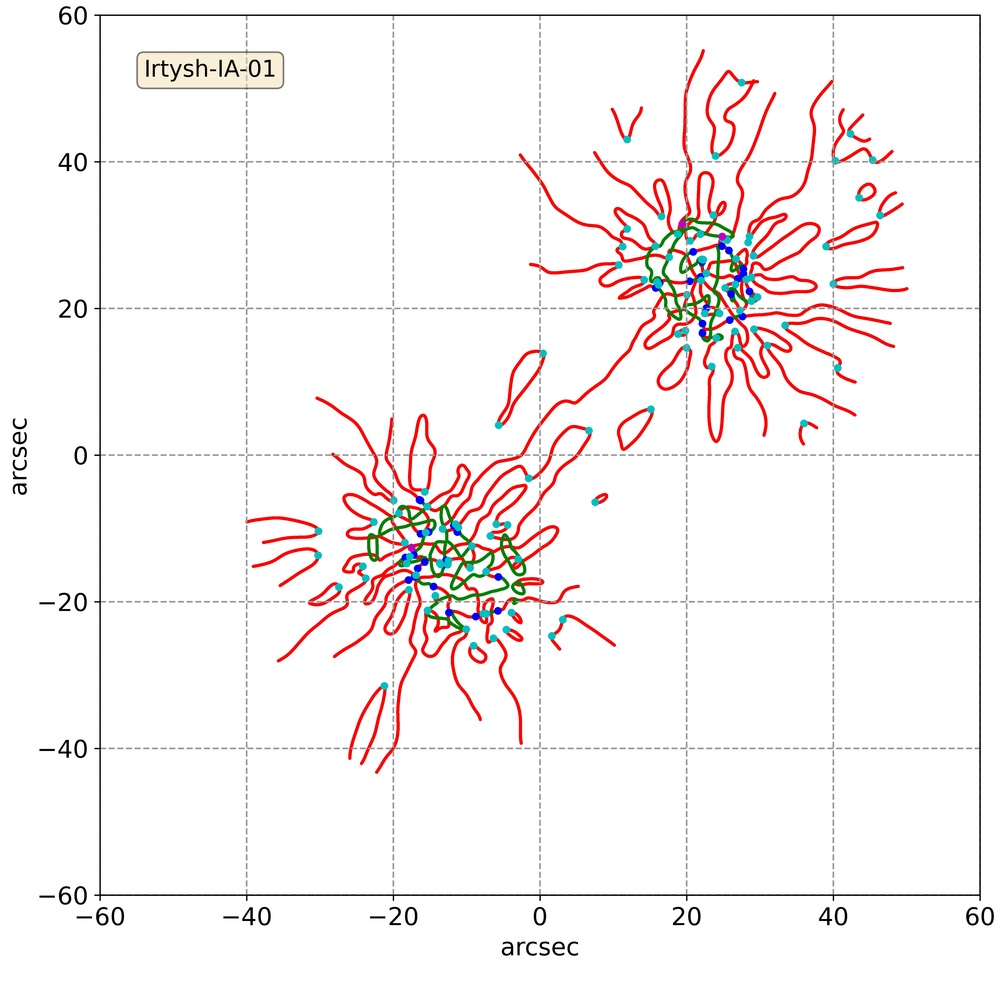}
  \includegraphics[width=\textwidth,height=7.4cm,width=7.5cm]{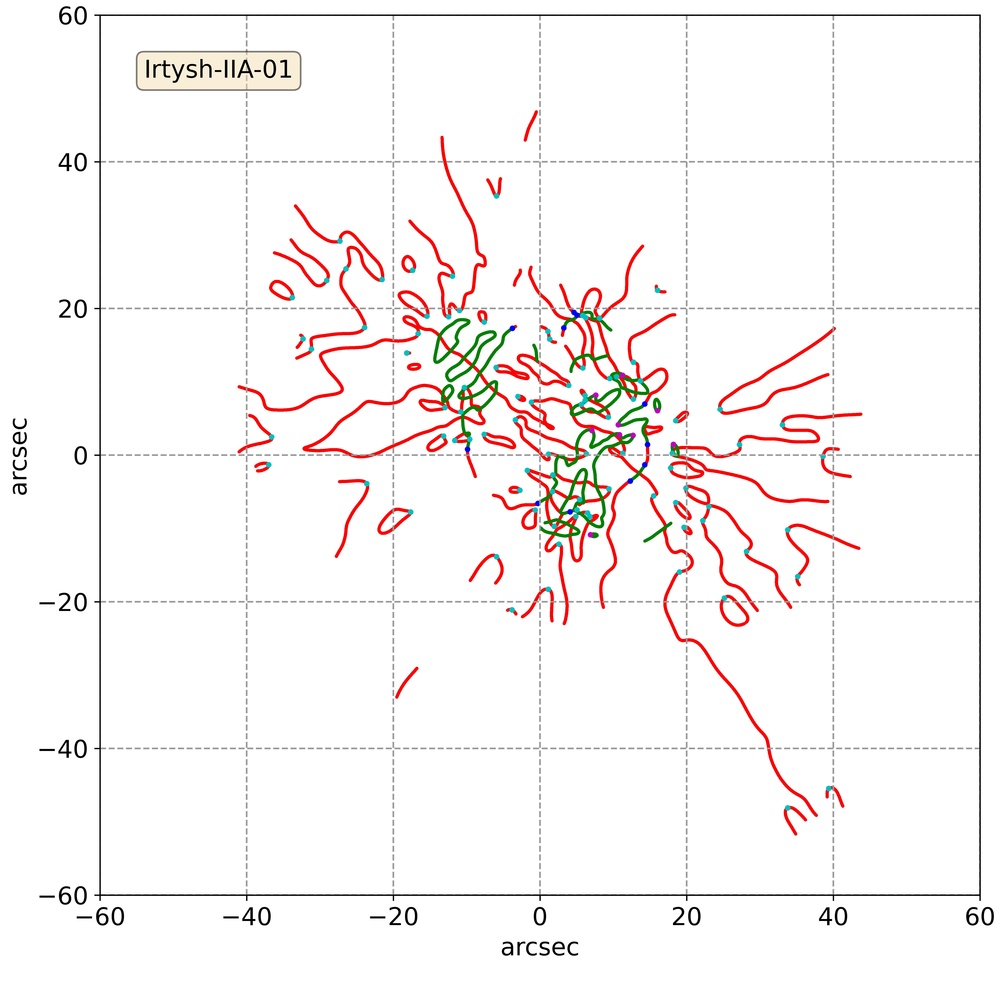}
  \includegraphics[width=\textwidth,height=7.4cm,width=7.5cm]{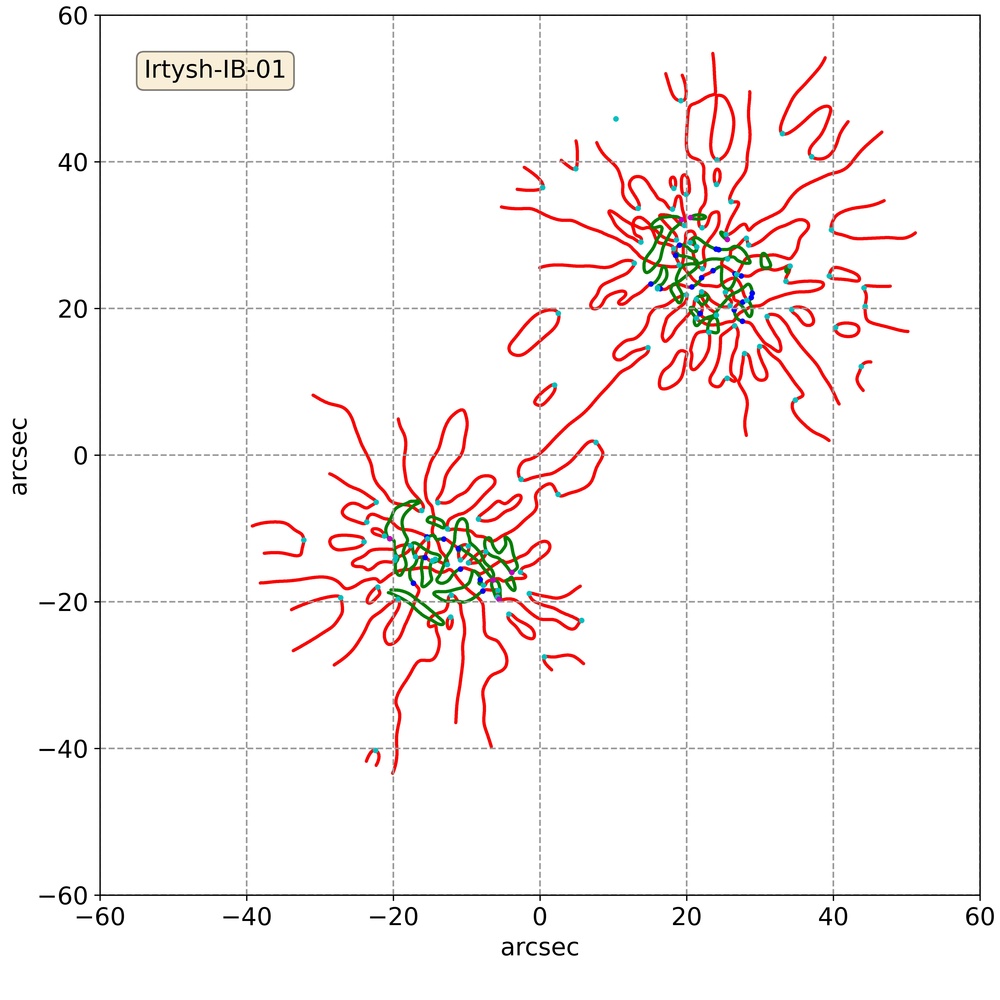}
  \includegraphics[width=\textwidth,height=7.4cm,width=7.5cm]{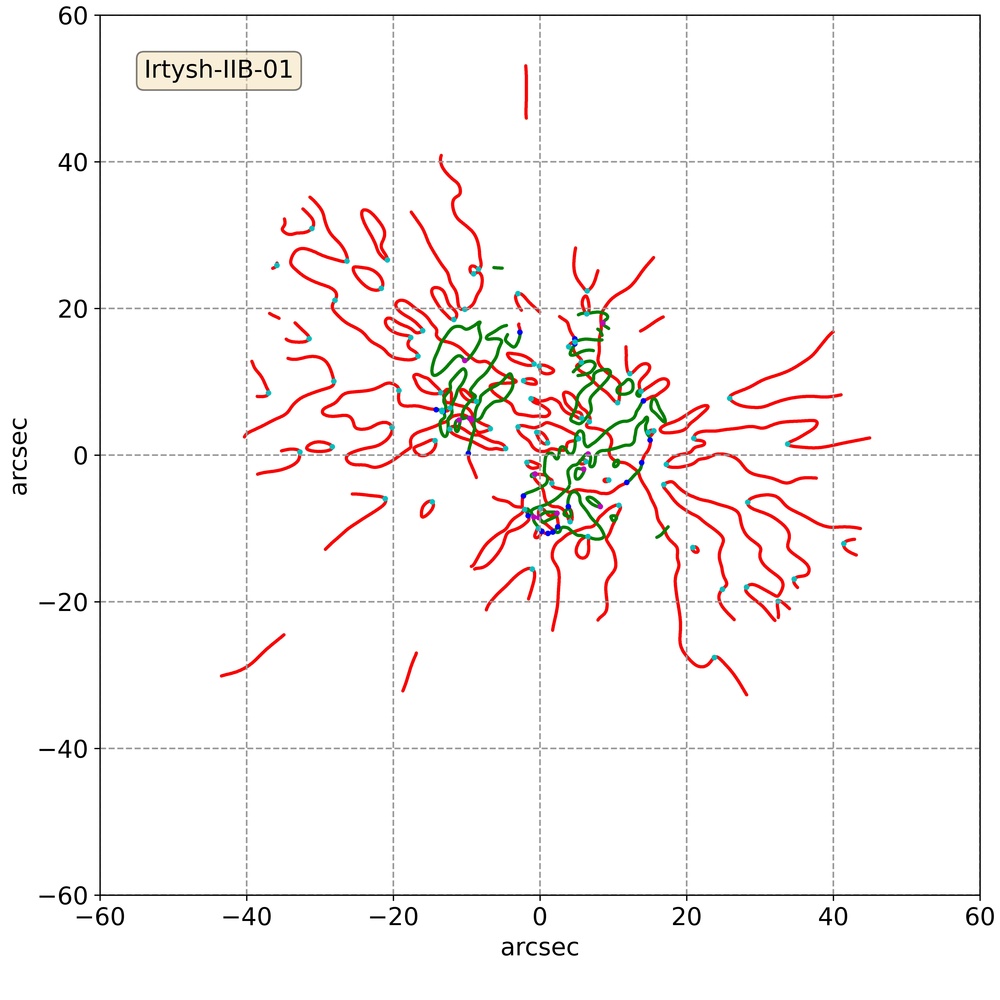}
  \includegraphics[width=\textwidth,height=7.4cm,width=7.5cm]{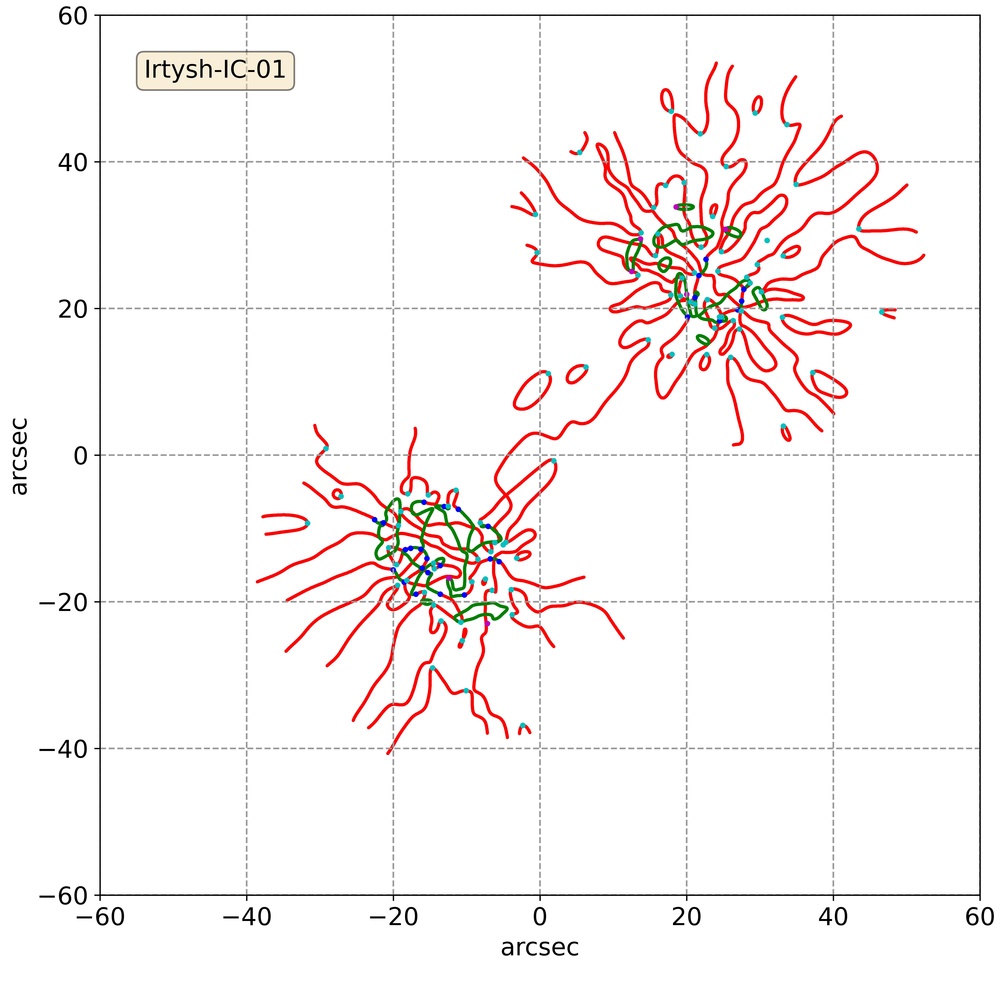}
  \includegraphics[width=\textwidth,height=7.4cm,width=7.5cm]{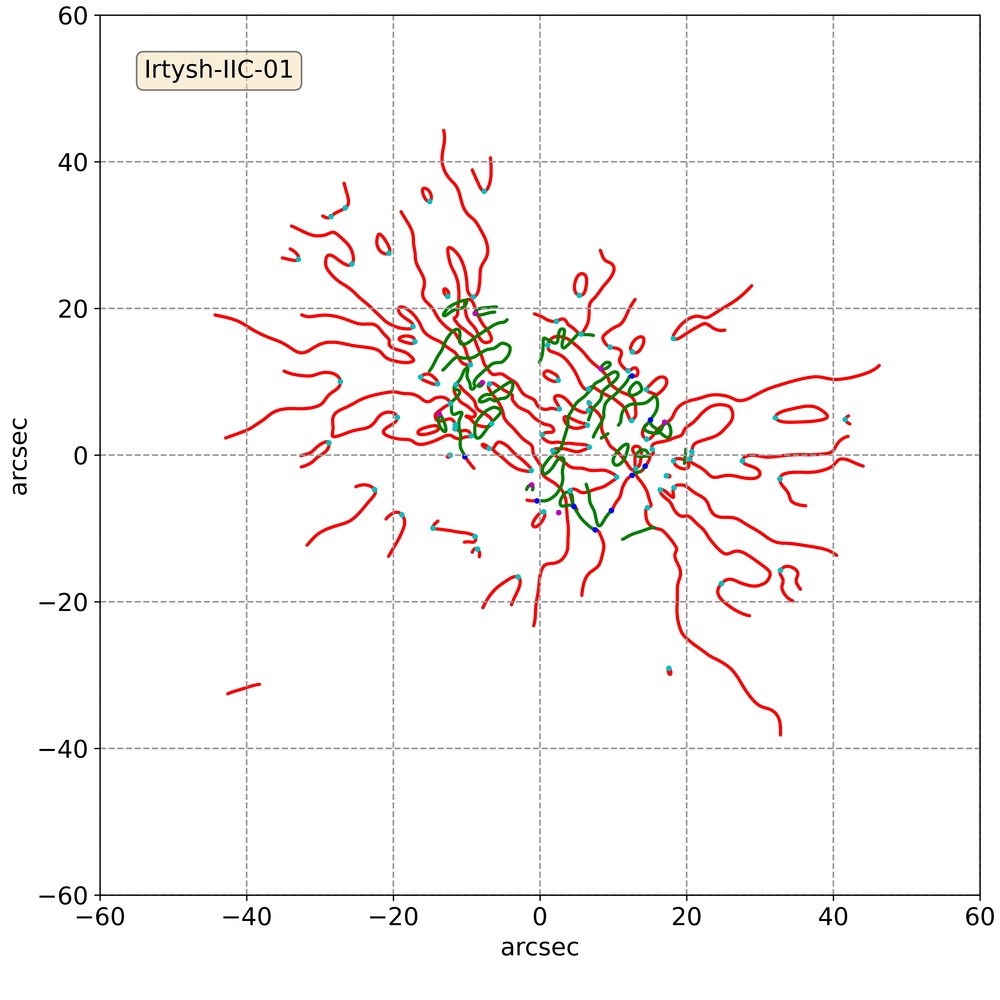}
  \caption{Singularity maps for individual runs of Irtysh clusters:
  Every panel represents one singularity map for one realization of different Irtysh
  reconstructions. The name of the realization is written with the 
  cluster name in the upper left corner. 
  The red and green lines represent the $A_3$-lines corresponding to the $\alpha$ 
  and $\beta$ eigenvalues of the deformation tensor. The blue points represent the
  (hyperbolic and elliptic) umbilics. The cyan and magenta points represent the
  swallowtail singularities corresponding to the $\alpha$ and $\beta$ eigenvalues 
  of the deformation tensor. }
  \label{fig: Simulated singularity indi}
\end{figure*}

\begin{figure*}
  \includegraphics[width=\textwidth,height=7.5cm,width=7.5cm]{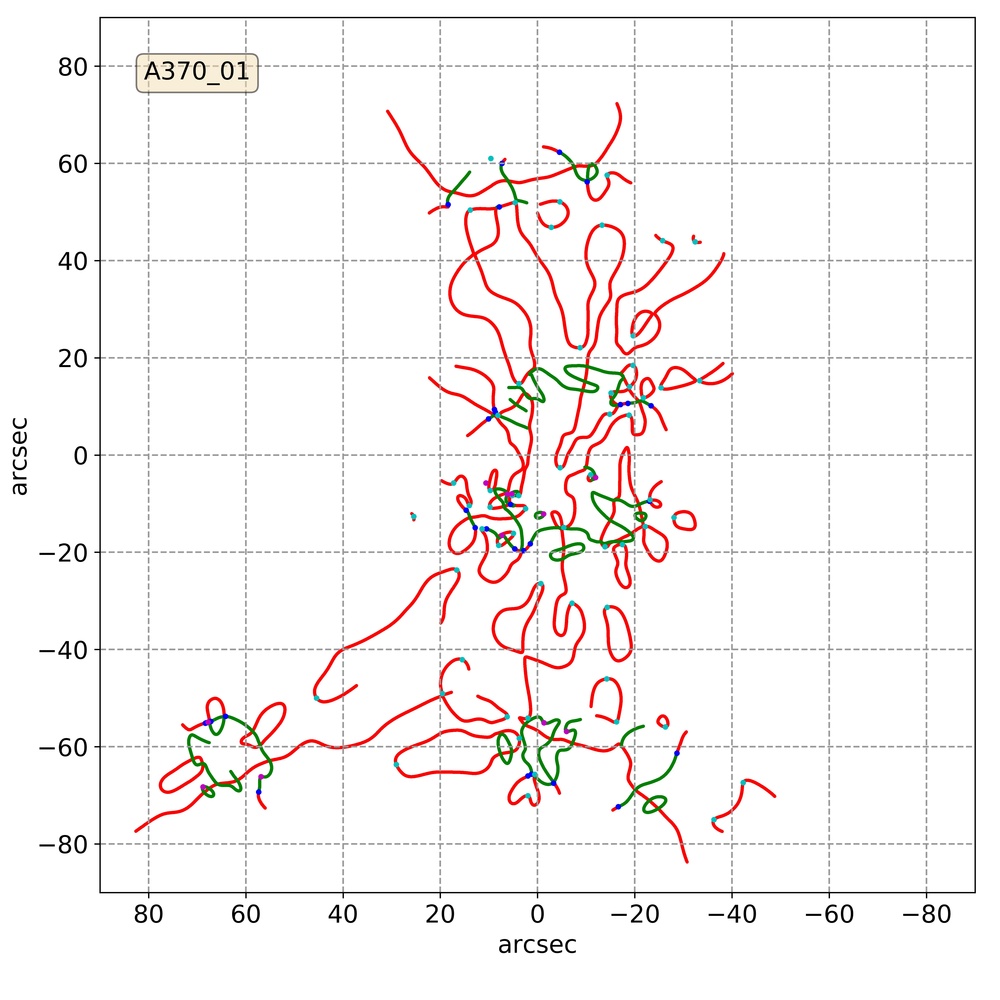}
  \includegraphics[width=\textwidth,height=7.5cm,width=7.5cm]{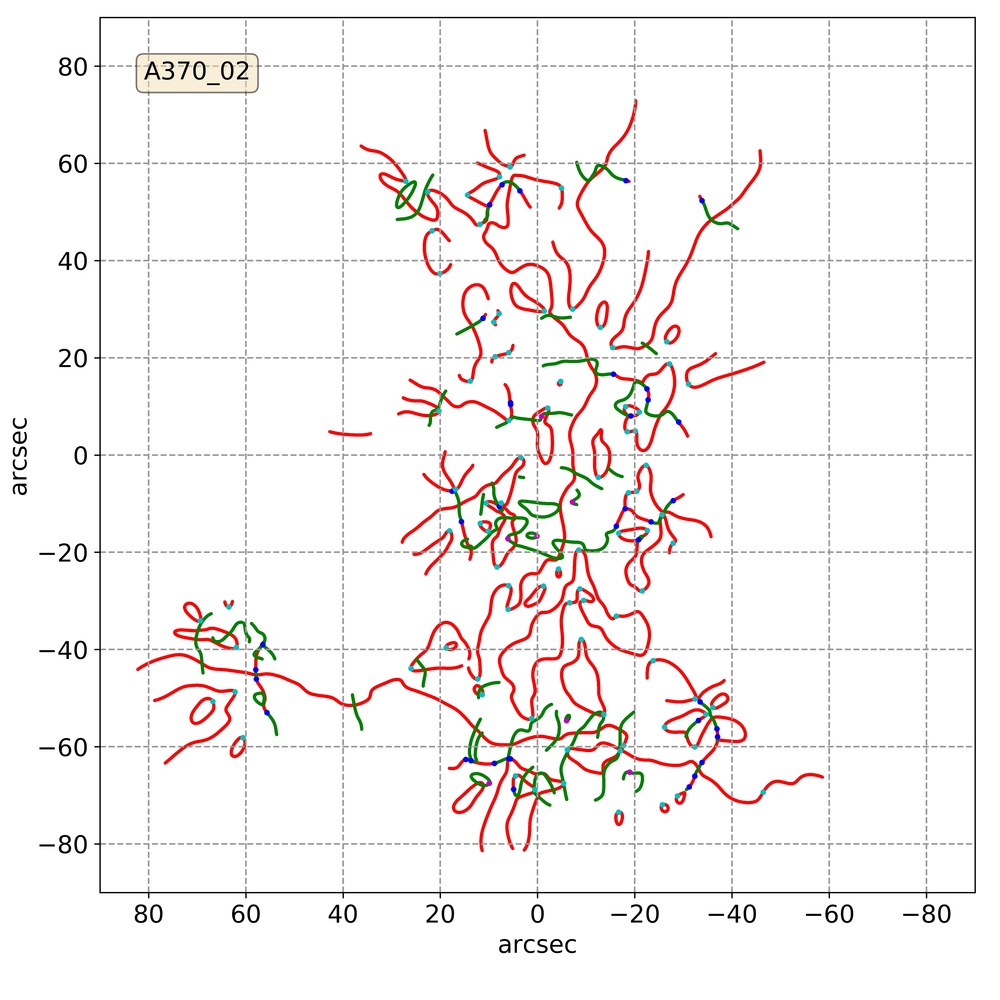}
  \includegraphics[width=\textwidth,height=7.5cm,width=7.5cm]{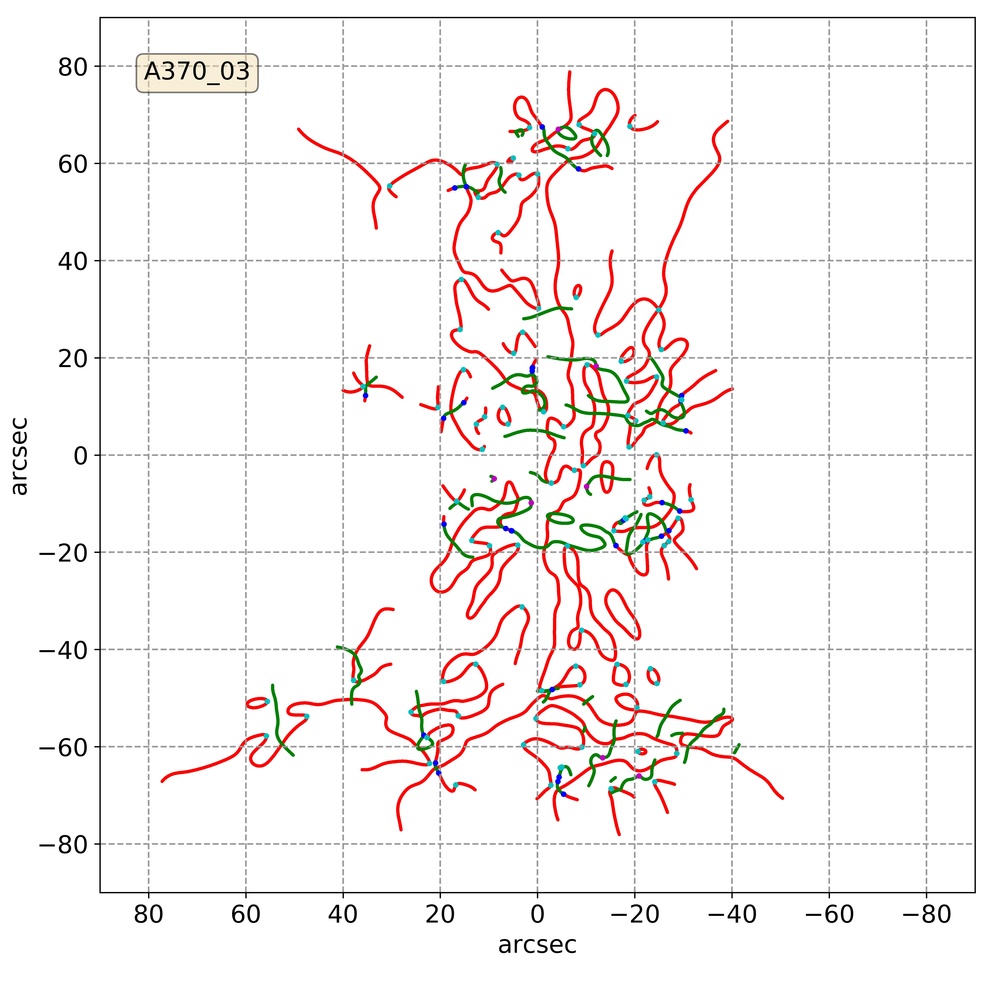}
  \includegraphics[width=\textwidth,height=7.5cm,width=7.5cm]{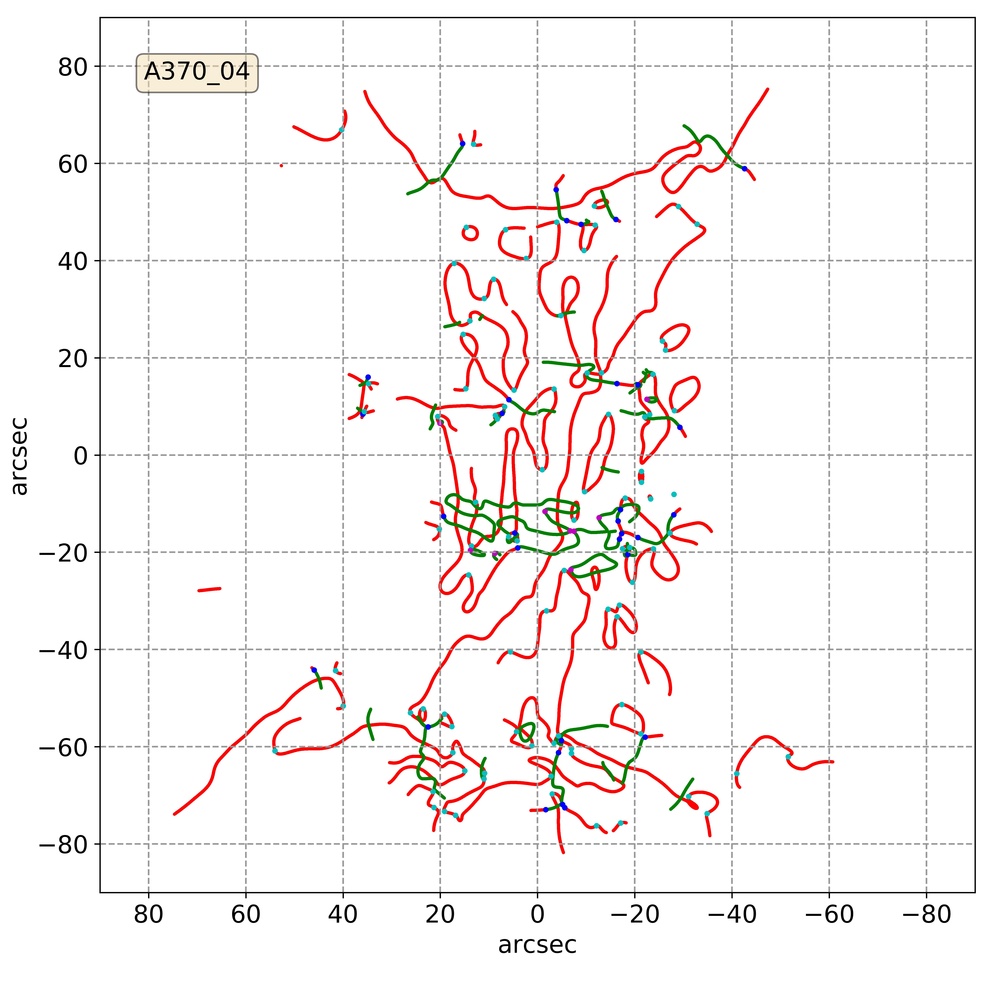}
  \includegraphics[width=\textwidth,height=7.5cm,width=7.5cm]{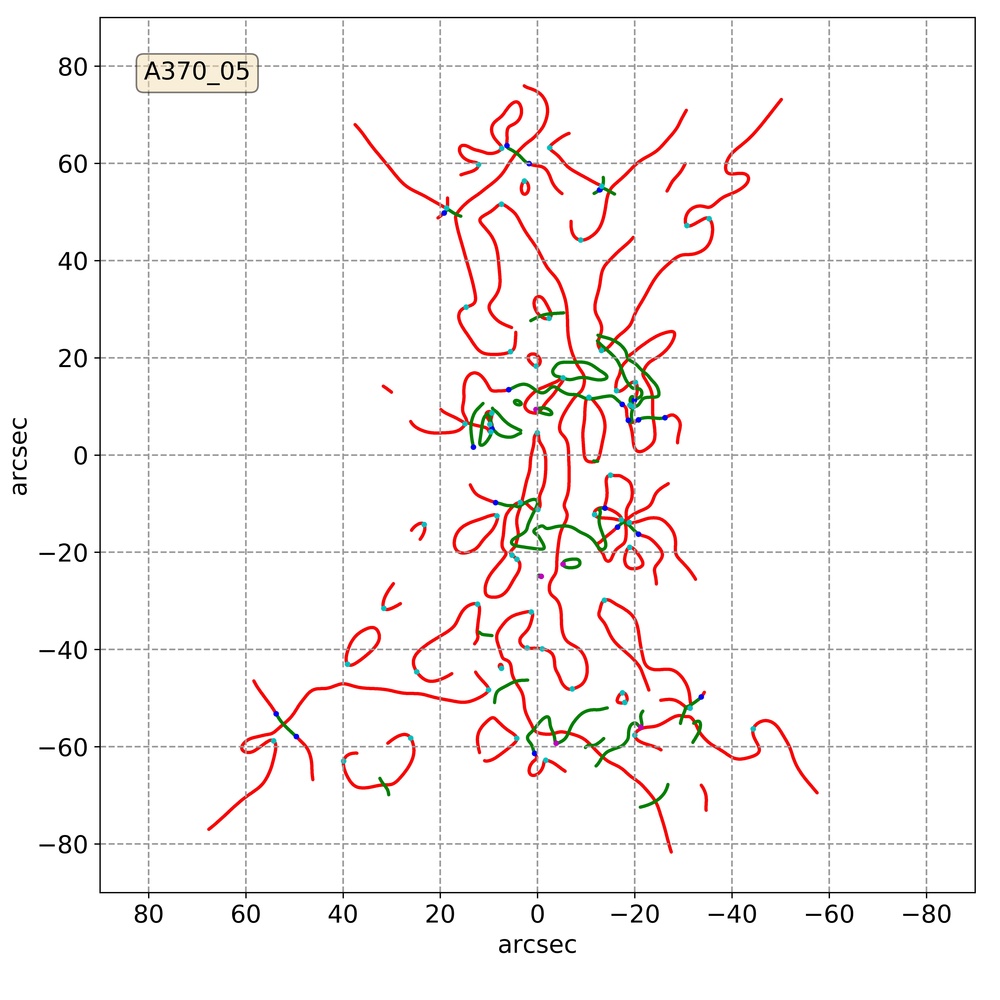}
  \includegraphics[width=\textwidth,height=7.5cm,width=7.5cm]{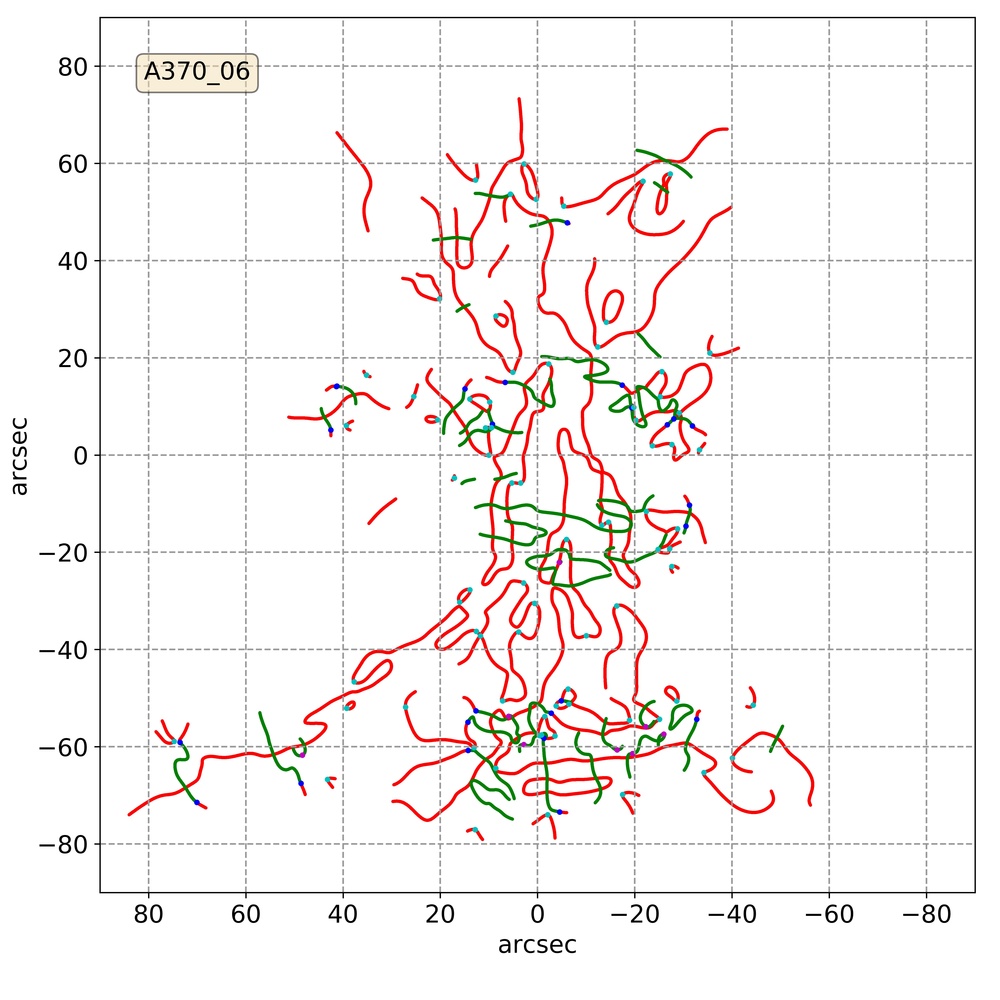}
  \caption{Singularity maps for individual runs of A370:
  Every panel represents one singularity map for one realization of A370. The name
  of the realization is written with the cluster name in the upper left corner. 
  The red and green lines represent the $A_3$-lines corresponding to the $\alpha$ 
  and $\beta$ eigenvalues of the deformation tensor. The blue points represent the
  (hyperbolic and elliptic) umbilics. The cyan and magenta points represent the
  swallowtail singularities corresponding to the $\alpha$ and $\beta$ eigenvalues 
  of the deformation tensor. }
  \label{fig: HFF singularity indi}
\end{figure*}

\bsp	
\label{lastpage}
\end{document}